\documentclass[AMA,12pt,STIX1COL]{WileyNJD-v2}

\usepackage{amsfonts,amsmath,amssymb,bm,xcolor,dirtytalk,natbib}
\usepackage{systeme}
\usepackage{econometrics}
\usepackage{graphics}
\usepackage{dsfont}
\usepackage{graphicx}
\usepackage{siunitx}
\usepackage[default]{lato}
\usepackage[T1]{fontenc}
\usepackage{cancel}
\usepackage{float}
\usepackage{booktabs}
\usepackage{pdfpages}
\usepackage{hyperref}
\usepackage{graphicx}
\usepackage{amsmath}
\usepackage{graphicx}
\usepackage{enumerate}
\usepackage{amsfonts}
\usepackage{physics,commath,derivative, algorithm, algpseudocode} 
\usepackage{url} 
\usepackage{enumitem}
\usepackage{subcaption}
\usepackage[margin=10mm]{caption}
\usepackage{comment}

\newcommand{\bz}{\mbox{\protect\boldmath $z$}}

\newcommand{\btheta}{{\bf\theta}}

\bmdefine\ttheta{\theta}

\newcommand{\bD}{\mbox{\protect\boldmath $D$}}

\newcommand{\bZ}{\mbox{\protect\boldmath $Z$}}
\newcommand{\bbeta}{{\bf \beta}}
\newcommand{\bomega}{{\bf \omega}}

\newcommand{\bGamma}{\mbox{\protect\boldmath $\Gamma$}}

\bmdefine\oomega{\omega}

\newcommand{\rmd}{\mathrm{d}}

\articletype{Research Article}%

\begin{document}


\title{Bayesian Federated Inference for Survival Models}
\author[1]{Hassan Pazira}
\author[2]{Emanuele Massa}
\author[3]{Jetty AM Weijers}
\author[2,4]{Anthony CC Coolen}
\author[1]{Marianne A Jonker}

\authormark{Pazira \textsc{et al}}

\address[1]{\orgdiv{Research Institute for Medical Innovation, Science department IQ Health, Research \& Education group Biostatistics}, \orgname{Radboud University Medical Center}, \orgaddress{\state{Nijmegen}, \country{Netherlands}}}

\address[2]{\orgdiv{Donders Institute, Faculty of Science}, \orgname{Radboud University}, \orgaddress{\state{Nijmegen}, \country{Netherlands}}}

\address[3]{\orgdiv{Department of Medical Oncology, Radboud Institute for Medical Innovation}, \orgname{Radboud university medical center}, \orgaddress{\state{Nijmegen}, \country{Netherlands}}}

\address[4]{\orgdiv{Saddle Point Science Europe, Mercator Science Park}, \orgaddress{\state{Nijmegen}, \country{Netherlands}}}

\corres{Marianne Jonker, Radboudumc Nijmegen, the Netherlands. \email{marianne.jonker@radboudumc.nl}}

\abstract{
In cancer research, overall survival and progression free survival are often analyzed with the Cox model. To estimate accurately the parameters in the model, sufficient data and, more importantly, sufficient events need to be observed. In practice, this is often a problem. Merging data sets from different medical centers may help, but this is not always possible due to strict privacy legislation and logistic difficulties. Recently, the Bayesian Federated Inference (BFI) strategy for generalized linear models was proposed. With this strategy the statistical analyses are performed in the local centers where the data were collected (or stored) and only the inference results are combined to a single estimated model; merging data is not necessary. The BFI methodology aims to compute from the separate inference results in the local centers what would have been obtained if the analysis had been based on the merged data sets. In this paper we generalize the BFI methodology as initially developed for generalized linear models to survival models. Simulation studies and real data analyses show excellent performance; i.e., the results obtained with the BFI methodology are very similar to the results obtained by analyzing the merged data.  An R package for doing the analyses is available.          

\bigskip

\noindent%
{\bf KEYWORDS:}\\ Cox regression, Federated Learning, multicenter data, non-shared data, rare cancer. 
}

\maketitle

\section{Introduction}
In cancer research, but also in many other research fields, overall survival (OS) is frequently regarded as the outcome of main interest; it is seen as the gold standard. For instance, the efficacy of a treatment is often evaluated based on overall survival and for cancer patient the remaining survival time is predicted based on the characteristics of the patient and the tumour. To have sufficient statistical power for concluding efficacy of a treatment or to estimate accurately a survival prediction model, sufficient events (i.e., deaths) need to be available in the data set. The latter may be a problem, especially if survival rates are reasonably high. This has led to a shift from overall survival as primary outcome to surrogate outcomes, like progression free survival and the objective response rate.\cite{Delgado} An alternative route would be to enlarge the data set and increase the total number of events (deaths) in the data by sharing data across different medical centers. Because of regulatory constraints and privacy legislation, this is not always possible, especially if the medical centers are located in different countries. 

Recently, Jonker et al\cite{Jonkera} proposed the so-called Bayesian Federated Inference (BFI) methodology to extract and combine information from different medical centers without actually sharing the data. In each center the local data are analyzed and specific inference results are sent to a central server, where the results from the separate medical centers are combined into one statistical model such that the estimated parameters are approximately equal to the estimates which would have been obtained if the data sets were pooled before the analysis was performed. With this methodology, privacy issues do not play any role as the patient data do not leave the hospital and can not be retrieved from the inference results that are sent to the central server. In Jonker et al\cite{Jonkera} the methodology has been proposed, explained and tested for generalized linear models (GLMs). The results were even better than expected; parameter estimates and outcome predictions obtained with the BFI methodology were very similar to those obtained by analyzing the combined data set (so pooling the data before doing the analysis). In this paper we generalize the BFI methodology to survival data.

The most popular model for analyzing survival data is the Cox proportional hazards (PH) model.\cite{Cox} Because this model does not belong to the family of GLMs, the methodology developed in Jonker at al is not directly applicable.\cite{Jonkera} The Cox PH model is a semi-parametric model. It consists of a regression part and an unknown baseline hazard function. For predicting the survival time of future patients, both the regression parameters and the baseline hazard function need to be estimated. Also when comparing the efficacy of treatments based on overall survival, an estimate of the hazard rate is of limited clinical value without an estimate of the baseline hazard to measure the actual gain in survival time. In the Cox model the baseline hazard function is left completely unconstrained (is non-parametric). It is unfeasible to obtain a non-parametric BFI estimate for this function without sharing patient survival data with the central server (where the estimates from the different centers are combined). Sharing patients’ survival data is often not allowed. Therefore, in this paper the BFI methodology is derived for some alternative, parametric estimators for the baseline hazard function. The parameter dimensionality of these estimators are low(er), but the patients’ privacy is guaranteed. Specifically, we consider several parametric baseline hazard functions of a fixed form (eponential, Weibull, Gompertz), an exponentiated polynomial hazard function, and the piecewise constant hazard function. Expressions for combining the local inference results are derived for each model and the methodology is tested. One of the motivations for this research was the estimation of overal survival models for salivary gland cancer, a rare cancer. As an application of the methodology we analyze data of patients with this cancer type.

Methodology for federated analyses is not new. Federated Learning (FL) is an iterative machine learning approach that was developed several years ago.\cite{McMahan} Its aim is the same as that of the BFI method: doing a combined analysis without sharing the data. However, the methods differ in implementation. With FL only the local data are used for training a machine learning system in each center. The locally estimated parameters are sent to the central server where the model is updated. The updated model is sent to the local centers again where the parameters are re-estimated based on the local data and the updated model. This procedure is “cycled” around the centers until convergence has been reached. The number of such iterations can easily exceed 100. As a comparison, for the BFI methodology a single estimate, so a single round, is sufficient. The FL strategy has evolved enormously in the last years \cite{Liu} and is even available for survival models.\cite{Lu, Rahman} It performs excellently in e.g., image analysis,\cite{Rieke,Gafni}, but can not easily handle statistical complexities like heterogeneity across populations, missing data, and complex statistical models. In contrast, with the BFI methodology the statistical models can be easily adjusted for heterogeneity and missing data.\cite{Jonkerb} 

The outline of the paper is as follows. In Section \ref{sec: BFI for survival} the methodology of the BFI for a general survival models is derived for a parametric baseline hazard function. Different choices for this baseline hazard function are described in Section \ref{sec: specific cases}. Next, in Section \ref{sec: sim studies} the results of multiple simulation studies are presented to study the performance of the BFI methodology for the different baseline hazard functions. A BFI analysis of data of salivary gland cancer patients is described in Section \ref{subsec:application}. The paper is summarized and discussed in Section \ref{sec:discussion}. In that section we also discuss some points for improvement which will be addressed in later projects. The BFI methodology for GLMs and survival models has been implemented in an R-package called BFI and an accompanying manual is available. \cite{Pazira}

\section{Bayesian Federated Inference for Survival models}
\label{sec: BFI for survival}
In this section we explain how to extend the Bayesian Federated Inference methodology as was originally proposed in Jonker at al\cite{Jonkera}, to models for time to event data. We consider the scenario in which data in $L$ medical centers are available, but these data can not be merged to a single data set. 

\subsection{Setting and Notation}
\label{subsec: notation}
Let, for patient $i$ from center $\ell$, $Y_{\ell i}$ be the time from a well specified time zero to an event of interest. The patient's vector of covariates measured at time zero is denoted as $\bZ_{\ell i}\in \mathbb{R}^p$. Suppose that $Y_{\ell i}| (\bZ_{\ell i}=\bz, \btheta)$ has density $y|\bz, \btheta \mapsto p(y|\bz, \btheta)$ and survival function $y|\bz, \btheta \mapsto S(y|\bz, \btheta)=\int_y^\infty  p(s|\bz, \btheta) \rmd{s}$, that are known up to the parameter $\btheta$ which itself has a density function $\btheta \mapsto \pi_\ell(\btheta)$. Let $C_{\ell i}$ be a non-negative right censoring time for this patient. We assume that conditional on $\bZ_{\ell i}$ (and the parameters), the censoring time $C_{\ell i}$ is independent of $Y_{\ell i}$. Further, let $T_{\ell i}$ be defined as $T_{\ell i} = \min\{Y_{\ell i}, C_{\ell i}\}$ and let $\Delta_{\ell i}=1\{T_{\ell i} \leq C_{\ell i}\}$ be the indicator that equals 1 if $T_{\ell i} \leq C_{\ell i}$ and 0 otherwise. There are $L$ data sets in $L$ centers which can not be merged. We assume that the patients' (stochastic) variables within the centers and across the centers are independent and identically distributed (conditional on the model parameters). 

For patient $i$ from center $\ell$ we observe the realisation $(t_{\ell i},\delta_{\ell i},\bz_{\ell i})$ of the stochastic triple $(T_{\ell i},\Delta_{\ell i},Z_{\ell i})$. The data set in center $\ell$ is denoted by $\bD_{\ell}$:
\begin{align*}
\bD_\ell = \{(t_{\ell 1}, \delta_{\ell 1}, \bz_{\ell 1}),\ldots, (t_{\ell n_\ell},\delta_{\ell n_\ell},\bz_{\ell n_\ell})\}, 
\end{align*}
where $n_\ell = | \bD_\ell|$ is the number of patients in data set $\bD_\ell$. The (fictive) combined data set, which is never actually created,  is denoted as $\bD = \cup_{\ell=1}^L \bD_\ell$ and contains data of  $n:=|\bD |=\sum_\ell^L n_\ell$ patients. 


Different survival models for the time to event variable $Y_{\ell i}$ correspond to different functional forms for $p(y|\bz,\btheta)$. The (semi-parametric) Cox proportional hazards model is arguably the most widely adopted, at least in medical applications.\cite{Cox} In this model the baseline hazard function is a priori unknown and unconstrained. Estimating the baseline hazard function is essential if the fitted model is to be used for predicting the survival times for new patients or interpreting estimated hazard rates. However, within the BFI framework (without sharing the survival times) this is only possible if a parametric form for baseline hazard function is assumed. For this reason, we will focus on parametric proportional hazards models.  

Conditional on the vector of covariates $\bZ_{\ell i}=\bz$ and $\btheta$, the hazard function for $Y_{\ell i}$ equals
\begin{align*}
y|\bz,\btheta \mapsto \lambda(y|\bz,\btheta) = \lambda_0(y|\bomega) \exp(\bz^\top \bbeta), \qquad \btheta = (\bbeta,\bomega)^{\top}     
\end{align*}
where $\lambda_0(.|\bomega)$ is the baseline hazard function which is known up to the parameter $\bomega\in \mathbb{R}^q$, and $\bbeta \in \mathbb{R}^p$ is the unknown vector with regression parameters. This yields the conditional density of the form
\begin{align*}
y|\bz,\btheta \mapsto    p(y|\bz,\btheta) = \lambda_0(y|\bomega)\exp\Big\{\bz^\top\bbeta-\Lambda_0(y|\bomega)\exp(\bz^\top\bbeta)\Big\}, 
\end{align*}
where $\Lambda_0(y|\bomega) = \int_{0}^{y} \lambda_0(s|\bomega)\rmd s$ is the cumulative baseline hazard function at time $y$. This class of parametric proportional hazards models gives, for data set $\bD_\ell$, the likelihood function
\begin{align*}
p(\mathbf{D}_{\ell}| \bbeta,\bomega) \propto \prod_{i=1}^{n_{\ell}}   \big\{\lambda_0(t_{\ell i }|\bomega)\exp(\bz_{\ell i }^\top\bbeta)\big\}^{\delta_{\ell i }}\exp\Big\{-\Lambda_0(t_{\ell i }|\bomega)\exp(\bz_{\ell i }^\top\bbeta)\Big\},
\end{align*}
where we left out the factors that depend on the  distribution functions for the censoring times and the covariates only and, thus, do not depend on the parameters of interest $\btheta = (\bbeta,\bomega)^\top$. 

\medskip

Suppose it had been possible to combine the data into the large data set $\bD$. For $\bD$, let $(\bbeta,\bomega)\mapsto \pi(\bbeta,\bomega)$ be the \emph{global} prior over the parameters of the model.
Then, its posterior density $(\bbeta,\bomega) \mapsto p(\bbeta,\bomega | \bD)$ would have been given by 
\begin{align*}
p( \bbeta,\bomega | \bD)  =\frac{p(\bD|\bbeta, \bomega)\pi(\bbeta,\bomega)}{Z (\bD)} =   \frac{\exp\{\Omega(\bbeta,\bomega| \bD)\}}{Z (\bD)},
\end{align*}
with $ Z(\bD) := \int p(\bD|\bbeta, \bomega)\pi(\bbeta,\bomega) \ \rmd \bbeta \rmd \bomega$ the normalizing constant of the posterior distribution and 
\begin{align}
\label{global_objective}
    \Omega(\bbeta,\bomega | \bD) =  \log \{\pi(\bbeta,\bomega)\} + \log \{p(\bD|\bbeta,\bomega)\}. 
\end{align}
The maximum a posteriori (MAP) estimator for $(\bbeta,\bomega)$, denoted as $(\widehat{\bbeta},\widehat{\bomega})$, is obtained by maximizing $(\bbeta,\bomega) \mapsto p( \bbeta,\bomega | \bD)$ with respect to $(\bbeta,\bomega)$: 
\begin{align*}
    (\widehat{\bbeta},\widehat{\bomega}) := \underset{(\bbeta,\bomega)}{\arg\max} \ p( \bbeta,\bomega | \bD)  = \underset{(\bbeta,\bomega)}{\arg\max} \  \Omega(\bbeta,\bomega| \bD).
\end{align*}
Data sharing, i.e., creating $\bD$, is often not possible, hence the above \emph{global} MAP estimator cannot be computed. Bayesian Federated Inference (BFI) seeks to obviate this issue by expressing the \emph{global} MAP estimator $(\widehat{\bbeta},\widehat{\bomega})$ in terms of the \emph{local} MAP estimators, i.e., the modes of the \emph{local} posteriors $(\bbeta,\bomega) \mapsto p_\ell(\bbeta,\bomega | \bD_\ell)$: 
\begin{align}
\label{local_post}
    (\widehat{\bbeta}_\ell,\widehat{\bomega}_\ell) := \underset{(\bbeta,\bomega)}{\arg\max} \ p_\ell( \bbeta,\bomega | \bD_\ell)  = \underset{(\bbeta,\bomega)}{\arg\max}  \ \Omega_\ell(\bbeta,\bomega| \bD_\ell) 
\end{align}
with
\begin{align*}
    \Omega_\ell(\bbeta,\bomega| \bD_\ell) = \log \{\pi_\ell(\bbeta,\bomega)\} + \log \{p(\bD_\ell|\bbeta,\bomega)\}  
\end{align*}
where $(\bbeta,\bomega)\mapsto\pi_\ell(\bbeta,\bomega)$ is the \emph{local} prior for the model parameters, and
\begin{align*}
      \log \{p(\bD_\ell|\bbeta,\bomega)\} = \sum_{i=1}^{n_\ell} \Big\{\delta_{\ell i} \big(\bz_{\ell i }^\top\bbeta + \log\{\lambda_0(t_{\ell i}|\bomega)\}\big) - \Lambda_0(t_{\ell i }|\bomega)\exp(\bz_{\ell i }^\top\bbeta)\Big\}. 
\end{align*}

In the next subsection the fictive global MAP estimator $(\widehat{\bbeta},\widehat{\bomega})$ is expressed in terms of the local MAP estimators $(\widehat{\bbeta}_\ell,\widehat{\bomega}_\ell), \ell=1,\ldots,L$. Then, once the local MAP estimates have been computed in the centers and sent to the central server, the global MAP estimates can be computed (approximated) via this expression.

\subsection{Deriving the BFI estimators}
\label{subsec: BFI}
By the definition of $\Omega(\bbeta,\bomega| \bD)$ in  (\ref{global_objective}) and the statistical independence between the data across the different centers,   $\Omega(\bbeta,\bomega| \bD)$ can be written as 
\begin{align}
\label{global_f_local}
    \Omega(\bbeta,\bomega| \bD) &=  \log \{\pi(\bbeta,\bomega)\} + \log \{p(\bD|\bbeta,\bomega)\}
    =  \log \{\pi(\bbeta,\bomega)\} +\sum_{\ell=1}^L \log \{p(\bD_\ell|\bbeta,\bomega)\} \nonumber\\
    &= \sum_{\ell=1}^L \Omega_\ell(\bbeta,\bomega| \bD_\ell)  + \log\{ \pi(\bbeta,\bomega)\} - \sum_{\ell=1}^L \log\{ \pi_{\ell}(\bbeta,\bomega) \}.
\end{align}

We approximate the local posteriors by a second order Taylor expansion around the local MAP estimators $(\widehat{\bbeta}_\ell,\widehat{\bomega}_\ell)$:
\begin{align}
    \label{local_laplace}
   \Omega_\ell(\bbeta,\bomega | \bD_\ell) 
    =\Omega_\ell(\widehat{\bbeta}_\ell,\widehat{\bomega}_\ell| \bD_\ell) - \frac{1}{2} 
    \begin{pmatrix}
        \bbeta - \widehat{\bbeta}_\ell\\
        \bomega - \widehat{\bomega}_\ell
    \end{pmatrix}^\top
    \widehat{\mathbf{M}}_\ell\begin{pmatrix}
        \bbeta - \widehat{\bbeta}_\ell\\
        \bomega - \widehat{\bomega}_\ell
    \end{pmatrix} + O_P\big(\|\bbeta-\widehat{\bbeta}_\ell\|^3 +\|\bomega-\widehat{\bomega}_\ell\|^3 \big) 
\end{align}
where $O_P(\|\bbeta-\widehat{\bbeta}_\ell\|^3 +\|\bomega-\widehat{\bomega}_\ell\|^3 ) =(\|\bbeta-\widehat{\bbeta}_\ell\|^3 +\|\bomega-\widehat{\bomega}_\ell\|^3)  O_p(1)$ with $O_p(1)$ a term that is bounded in probability for the sample size going to infinity\cite{Vaart} and the matrix $\widehat{\mathbf{M}}_\ell$ is equal to minus the Hessian matrix of $(\bbeta,\bomega)\mapsto \Omega_\ell(\bbeta,\bomega | \bD_\ell)$ evaluated at the MAP estimator. The linear term is missing in the expansion as this term equals zero by the definition of the MAP estimator.

We assume zero mean Gaussian priors in the combined and local data sets, with inverse covariance matrices $\bGamma$ and $\bGamma_\ell$:
\begin{align*}
\pi(\bbeta,\bomega) &= \bigg(\frac{\det (\bGamma)}{(2\pi)^d}\bigg)^{\frac{1}{2}}\exp\Bigg\{-\frac{1}{2}\begin{pmatrix}
        \bbeta\\
        \bomega
    \end{pmatrix}^\top \bGamma \begin{pmatrix}
        \bbeta\\
        \bomega
    \end{pmatrix}\Bigg\},\\[4pt]
    \pi_\ell(\bbeta,\bomega) &= \bigg(\frac{\det (\bGamma_\ell)}{(2\pi)^d}\bigg)^{\frac{1}{2}}\exp\Bigg\{-\frac{1}{2}\begin{pmatrix}
        \bbeta\\
        \bomega
    \end{pmatrix}^\top \bGamma_\ell \begin{pmatrix}
        \bbeta\\
        \bomega
    \end{pmatrix}\Bigg\}.
\end{align*} 
By this Gaussian prior assumption the matrix $\widehat{\mathbf{M}}_\ell$ can be decomposed into $ \widehat{\mathbf{M}}_\ell := \widehat{\mathbf{I}}_\ell + \bGamma_\ell$,  where $\widehat{\mathbf{I}}_\ell:= \mathbf{I}_\ell(\widehat{\bbeta}_\ell,\widehat{\bomega}_\ell)$ is defined as 
\begin{align*}
   \mathbf{I}_\ell(\bbeta,\bomega) = 
    \begin{bmatrix}
        \mathbf{I}_\ell^{\bbeta \bbeta}(\bbeta,\bomega) &~~~ \mathbf{I}_\ell^{\bbeta \bomega}(\bbeta,\bomega)\\
        \mathbf{I}_\ell^{\bomega \bbeta}(\bbeta,\bomega) &~~~ \mathbf{I}_\ell^{\bomega \bomega}(\bbeta,\bomega)
    \end{bmatrix}
\end{align*}
with the blocks given by 
\begin{align*}
    \mathbf{I}_\ell^{\bbeta \bbeta}(\bbeta,\bomega) &= \sum_{i=1}^{n_\ell}\Big\{ \bz_{\ell i }\bz_{\ell i }^{\top} \Lambda_0(t_{\ell i }|\bomega)\exp(\bz_{\ell i }^\top\bbeta) \Big\}\\
    \mathbf{I}_\ell^{\bbeta \bomega}(\bbeta,\bomega) &= \mathbf{I}_\ell^{\bomega \bbeta}(\bbeta,\bomega)= \sum_{i=1}^{n_\ell} \Big\{\bz_{\ell i } \nabla_{\bomega}\Lambda_0(t_{\ell i }|\bomega)^{\top} \exp(\bz_{\ell i }^\top\bbeta)\Big\} \\
    \mathbf{I}_\ell^{\bomega \bomega}(\bbeta,\bomega) &= \sum_{i=1}^{n_\ell} \Big\{\nabla_{\bomega}^2 \Lambda_0(t_{\ell i }|\bomega)\exp(\bz_{\ell i }^\top\bbeta) - \delta_{\ell i}\nabla_{\bomega}^2\log\{\lambda_0(t_{\ell i}|\bomega)\}\Big\}.
\end{align*}
The derivatives with respect to $\bomega$ can be computed explicitly once the parametric form of the baseline function has been chosen (see Section \ref{sec: specific cases}). Inserting the Taylor approximation given in (\ref{local_laplace}) and the Gaussian prior densities into the expression of $\Omega(\bbeta,\bomega|\bD)$ in (\ref{global_f_local}), gives
\begin{align}
\label{BFI_objective}
    \Omega(\bbeta,\bomega| \bD) = \Omega_{BFI}(\bbeta,\bomega) + O_P\Big(\underset{\ell=1,\dots,L}{\max}\big\{\|\bbeta-\widehat{\bbeta}_\ell\|^3 +\|\bomega-\widehat{\bomega}_\ell\|^3\big\} \Big) 
\end{align}
where the BFI surrogate objective function $(\bbeta,\bomega) \mapsto \Omega_{BFI}(\bbeta,\bomega)$ is defined as 
\begin{align*}
    \lefteqn{\Omega_{BFI}(\bbeta,\bomega):=}\\ 
    &~~~~ \sum_{\ell=1}^L \Bigg\{  \Omega_\ell(\widehat{\bbeta}_\ell,\widehat{\bomega}_\ell| \bD_\ell) - \frac{1}{2} 
    \begin{pmatrix}
        \bbeta - \widehat{\bbeta}_\ell\\
        \bomega - \widehat{\bomega}_\ell
    \end{pmatrix}^\top
    \widehat{\mathbf{M}}_\ell\begin{pmatrix}
        \bbeta - \widehat{\bbeta}_\ell\\
        \bomega - \widehat{\bomega}_\ell
    \end{pmatrix} \Bigg\} - \frac{1}{2} \begin{pmatrix}
        \bbeta\\
        \bomega 
    \end{pmatrix}^\top \big(\bGamma- \sum_{\ell=1}^L \bGamma_\ell\big) \begin{pmatrix}
        \bbeta\\
        \bomega 
    \end{pmatrix} \nonumber \\[6pt]
     &= \sum_{\ell=1}^L \Bigg\{  \Omega_\ell(\widehat{\bbeta}_\ell,\widehat{\bomega}_\ell| \bD_\ell) - \frac{1}{2} 
    \begin{pmatrix}
        \widehat{\bbeta}_\ell\\
        \widehat{\bomega}_\ell
    \end{pmatrix}^\top
    \widehat{\mathbf{M}}_\ell\begin{pmatrix}
        \widehat{\bbeta}_\ell\\
        \widehat{\bomega}_\ell
    \end{pmatrix}  + 
    \begin{pmatrix}
        \bbeta \\
        \bomega 
    \end{pmatrix}^\top
    \widehat{\mathbf{M}}_\ell\begin{pmatrix}
        \widehat{\bbeta}_\ell\\
        \widehat{\bomega}_\ell
    \end{pmatrix} \Bigg\}  - \frac{1}{2} \begin{pmatrix}
        \bbeta\\
        \bomega 
    \end{pmatrix}^\top \big(\bGamma + \sum_{\ell=1}^L \big\{  \widehat{\mathbf{M}}_\ell - \bGamma_\ell \big\} \big) \begin{pmatrix}
        \bbeta\\
        \bomega 
    \end{pmatrix}.\\
\end{align*}

The function $(\bbeta,\bomega) \mapsto \Omega_{BFI}(\bbeta,\bomega)$ is a quadratic function with respect to $\bbeta$ and $\bomega$ and can be easily maximized with respect to $(\bbeta,\bomega)$. The value in which $\Omega_{BFI}(\bbeta,\bomega)$ attains its maximum is  denoted as $(\widehat{\bbeta}_{BFI},\widehat{\bomega}_{BFI})$, and is given by 
\begin{align}
\label{est_bfi}
 \begin{pmatrix}
 \widehat{\bbeta}_{BFI} \\ 
 \widehat{\bomega}_{BFI}
\end{pmatrix} &=  \widehat{\mathbf{M}}_{BFI}^{-1} \sum_{\ell=1}^{L} \widehat{\mathbf{M}}_{\ell}  \begin{pmatrix}
 \widehat{\bbeta}_{\ell} \\ 
 \widehat{\bomega}_{\ell}
\end{pmatrix} = \big(\bGamma + \sum_{\ell=1}^L \widehat{\mathbf{I}}_\ell \big)^{-1} \sum_{\ell=1}^{L}   \big(\widehat{\mathbf{I}}_\ell + \bGamma_\ell\big)\begin{pmatrix}
 \widehat{\bbeta}_{\ell} \\ 
 \widehat{\bomega}_{\ell}
\end{pmatrix} ,\\[6pt]
\label{M_bfi}
\widehat{\mathbf{M}}_{BFI}  &=   \mathbf{\bGamma} +\sum_{\ell=1}^{L} \big(\widehat{\mathbf{M}}_{\ell} - \bGamma_{\ell}\big) =  \sum_{\ell=1}^{L}  \widehat{\mathbf{I}}_\ell +  \mathbf{\bGamma}\ .
\end{align}

For $\widehat{\bbeta}_\ell$ and $\widehat{\bomega}_\ell$ in a small neighborhood of $(\widehat{\bbeta}_{BFI},\widehat{\bomega}_{BFI})$ for every $\ell$, the remainder term in (\ref{BFI_objective}) will be small compared to $\Omega_{BFI}(\widehat{\bbeta}_{BFI},\widehat{\bomega}_{BFI})$. If, moreover, this remainder term behaves well in the sense that it is of bounded variation, the estimator $(\widehat{\bbeta}_{BFI},\widehat{\bomega}_{BFI})$ will be close to global MAP estimators in the combined data set: $(\widehat{\bbeta},\widehat{\bomega})$.

If the local sample sizes are sufficiently large, the MAP estimates and the maximum likelihood estimates will be very similar and the matrices $\widehat{\mathbf{I}}_\ell$ in (\ref{M_bfi}) are close to their local Fisher information matrices and, thus, positive definite. Since, moreover, the matrix $\bGamma$ is positive definite by definition,  $\widehat{\mathbf{M}}_{BFI}$ will be positive definite as well and the estimates in (\ref{est_bfi}) and (\ref{M_bfi}) are well defined. However, even if in a single center $\widehat{\mathbf{I}}_\ell$ is not positive definite (which would be unexpected), the matrix $\widehat{\mathbf{M}}_{BFI}$ may still be positive definite. 

We stress that the BFI estimators in (\ref{est_bfi}) and (\ref{M_bfi}) can be computed via linear operations in a single communication round from the local centers to the global central server \emph{without} sharing the data, but only the local MAP estimates $(\widehat{\bbeta}_\ell,\widehat{\bomega}_\ell,\widehat{\mathbf{M}}_\ell)$. An iterative algorithm to obtain estimators, as would have been required in Federated Learning procedures, is here not necessary.

The (fictive) MAP estimator $(\widehat{\bbeta},\widehat{\bomega})$ (computed based on the combined data set) is approximately Gaussian with mean $(\widehat{\bbeta},\widehat{\bomega})$ and inverse covariance matrix $\widehat{\mathbf{M}}$.\cite{Vaart} With the BFI methodology $(\widehat{\bbeta},\widehat{\bomega})$ and $\widehat{\mathbf{M}}$ are approximated by the BFI-estimators $(\widehat{\bbeta}_{BFI},\widehat{\bomega}_{BFI})$ and $\widehat{\mathbf{M}}_{BFI}$ and thus  $(\widehat{\bbeta},\widehat{\bomega})$ is approximately Gaussian with mean $(\widehat{\bbeta}_{BFI},\widehat{\bomega}_{BFI})$ and inverse covariance matrix $\widehat{\mathbf{M}}_{BFI}$. Now, using the normality yields approximated credible intervals for the parameters: for the $k^{th}$ element of $(\bbeta,\bomega)$ its approximate $(1-2\alpha) 100\%$ credible interval equals $(\widehat{\bbeta}_{BFI},\widehat{\bomega}_{BFI})_k \pm \xi_\alpha \; (\widehat{\mathbf{M}}_{BFI}^{-1})_{k,k}^{1/2},$ for $\xi_\alpha$ the upper $\alpha$-quantile of the standard Gaussian distribution.

\section{Choices for the baseline hazard function}
\label{sec: specific cases}
In the previous section we derived the BFI estimators for a parametric survival model in which the baseline hazard function has a general parametric form. In this section we consider different choices for this form. In Subsection \ref{subsec: simple_par} we start with the Weibull and Gompertz parameterizations of the baseline hazard, which depend on only two parameters. We then continue with more complex models, where the number of parameters can, in principle, be increased arbitrarily, allowing for more flexibility (i.e, as the number of parameters increases, the model can more accurately represent the shape of the underlying true function). In particular in Subsection \ref{subsubsec:piecewise_exp} we consider the simple, yet quite flexible, Piece-Wise Exponential model, where, as the name suggests, the baseline hazard function is approximated with a piece-wise constant function. Finally, Subsection \ref{subsec:taylor} deals with the Exponentiated Polynomial parameterization, where the logarithm of the baseline hazard function is assumed to be a polynomial. 

If categorical variables are included in the model, one of the categories is seen as reference category, and for the other categories a dummy-variable is defined. In most regression models the effect of the reference group is hidden in the intercept. In the semi-parametric Cox model no intercept is included as this intercept is incorporated in the baseline hazard function. If the baseline hazard function is parametric, the intercept need to be included  in one of the parameters of the baseline hazard function. From the functional form of the baseline hazard functions given below, it is clear that this happens.

\subsection{Simple Parametric Proportional Hazards Models}
\label{subsec: simple_par}
The models discussed in this subsection are parameterized by a scale parameter $ \exp(\omega_1) \in \mathbb{R}^+$ where $\omega_1 \in \mathbb{R}$ and a location parameter $\exp(\omega_2) \in \mathbb{R}^+$ with $\omega_2 \in \mathbb{R}$. Specifically, we examine the Weibull and Gompertz parameterizations of the baseline hazard rate. \cite{Kalbfleisch_2011} These are well-known simple parametric models that assume fixed functional forms for the baseline hazard function.

In the Weibull model the baseline hazard function and the cumulatives baseline hazard functions are defined by 
\begin{align*}
\lambda_0(t|\omega_1,\omega_2)=\exp(\omega_1 + \omega_2) \, t^{\exp(\omega_2)-1} \quad \text{and} \quad \Lambda_0(t|\omega_1,\omega_2)=\exp(\omega_1) \, t^{\exp(\omega_2)}.
\end{align*}
For computing the matrix $\widehat{M}_\ell$ the first and second derivatives of the cumulative baseline hazard function with respect to the parameters need to be computed. These derivatives equal  
\begin{align*}
\nabla_{\bomega}\Lambda_0(t|\bomega)=\begin{bmatrix} \Lambda_0(t|\bomega)\\ t \log(t) \, 
 \lambda_0(t|\bomega)
\end{bmatrix}, \quad \nabla_{\bomega}^2\Lambda_0(t|\bomega)=\begin{bmatrix}
 \Lambda_0(t|\bomega) &~ t \log(t) \, 
 \lambda_0(t|\bomega)\\
        t \log(t) \, 
 \lambda_0(t|\bomega) &~~ t \log(t) \, 
 \lambda_0(t|\bomega) (1+\log(t) \exp(\omega_2))
    \end{bmatrix},
\end{align*}
and 
\begin{align*}
\nabla_{\bomega}^2 \log \{\lambda_0(t|\bomega)\}=\begin{bmatrix}
        0 & 0\\
        0 &~ \log(t) \exp(\omega_2)
    \end{bmatrix}.
\end{align*}

\medskip

In the Gompertz model the baseline hazard function and the cumulative baseline hazard function are equal to 
\begin{align*}
\lambda_0(t|\omega_1,\omega_2)=\exp(\omega_1 + \exp(\omega_2) \, t) \quad \text{and} \quad  \Lambda_0(t|\omega_1,\omega_2)= \exp(\omega_1 - \omega_2) \, \big(\exp(\exp(\omega_2) \, t) -1 \big).
\end{align*}
The gradients equal
\begin{align*}
\nabla_{\bomega}\Lambda_0(t|\bomega)=
\begin{bmatrix} \Lambda_0(t|\bomega) \\ t \, \lambda_0(t|\bomega) - \Lambda_0(t|\bomega) \end{bmatrix}, \quad  \nabla_{\bomega}^2 \Lambda_0(t|\bomega)=\begin{bmatrix}
        \Lambda_0(t|\bomega) &~ t \, \lambda_0(t|\bomega) - \Lambda_0(t|\bomega) \\
        t \, \lambda_0(t|\bomega) - \Lambda_0(t|\bomega) &~~~ t \, \lambda_0(t|\bomega) \, (t \, \exp(\omega_2) - 1) + \Lambda_0(t|\bomega)
    \end{bmatrix},
\end{align*}
and 
\begin{align*}
\nabla_{\bomega}^2 \log \{\lambda_0(t|\bomega)\}=\begin{bmatrix}
        0 & 0\\
        0 &~ t \, \exp(\omega_2)
    \end{bmatrix}.
\end{align*}

Both Weibull and Gompertz parameterizations of the baseline hazard function $\lambda_0(\cdot)$ are monotonic: the Gompertz baseline hazard function is increasing, while the Weibull baseline hazard function might be increasing ($\omega_2>1$) or decreasing ($\omega_2<1$).

\subsection{Piece-wise Exponential Model}
\label{subsubsec:piecewise_exp}
The introduction of the piece-wise exponential baseline hazard function dates back to the work of Breslow \cite{Breslow_72, Breslow_75} in the 70s, followed by several studies on statistical inference\cite{Kalbfleisch_73, Friedman_82} in the following decade. The baseline hazard function is approximated by a function which is constant on $q$ intervals delimited by $q+1$ end points $\tau_0,\dots,\tau_{q}$. The basis functions are defined as
\begin{align}
    \mathbf{b}(t) = \big(b_0(t), \dots, b_{q-1}(t)\big)^\top, \quad b_k(t) := \bm{I}[\tau_k < t<\tau_{k+1}], \quad k = 0, \dots, q-1,
\end{align}
where $\bm{I}[\tau_k < t<\tau_{k+1}]$ equals 1 if $\tau_k < t<\tau_{k+1}$ and 0 otherwise. For these basis functions
the baseline hazard function can be written as 
\begin{align*}
    \lambda_0(t|\bomega) = \exp(\bomega)^\top\mathbf{b}(t)
\end{align*}
with $\exp(\bomega) = (\exp\normalsize(\omega_0\normalsize),\dots, \exp\normalsize(\omega_{q-1}\normalsize))^\top$.
This gives the cumulative hazard function
\begin{align*}
    \Lambda_0(t|\bomega) = \exp(\bomega)^\top\mathbf{B}(t), 
\end{align*}
where 
\begin{align*}
     \mathbf{B}(t) = \big(B_0(t), \dots, B_{q-1}(t)\big)^\top, \quad  B_k(t) := \bm{I}[t > \tau_k]\min\{t-\tau_k,\tau_{k+1}-\tau_k\}.
\end{align*}
The gradient of the cumulative hazard function equals 
\begin{align*}   \nabla_{\bomega}\Lambda_0(t|\bomega) = \exp(\bomega)\circ \mathbf{B}(t)
\end{align*}
where $\circ$ indicates  the component-wise product.
Further, $\nabla_{\bomega}^2 \log \{\lambda_0(t|\bomega)\} = \bm{0}_{q\times q}$, a $q\times q$ matrix with zeroes,  and 
\begin{align}
    \nabla_{\bomega}^2 \Lambda_0(t|\bomega) = {\rm diag}\big(\exp(\bomega)\circ \mathbf{B}(t)\big) 
\end{align}
where the ${\rm diag}$ operator returns a square diagonal matrix with the input vector on the diagonal.

\medskip

The piece-wise parametrization can be regarded as a relatively simple fully parametric version of the Cox model. 
Its advantage over the parametric models of the previous subsection is its flexibility. In fact the piece-wise exponential model is a special case of a B-spline parameterization of the baseline hazard function. \cite{Lazaro} The choice of the knots (i.e., $\tau_1, \dots, \tau_{q-1}$)  plays a central role: \say{well chosen} knots will lead to a better fit of the data. Although several recipes have been put forward, there is no unique agreed methodology to chose the position of the knots. It has been noticed in literature \cite{Harrell_2001} that the positions of the knots do not generally impact on the quality of the fit: the number of knots is crucial. Within the BFI framework it is essential that the same knots are used in every center. Practically it might well be that the researchers across the different centers agree on a reasonable placement of the knots. Alternatively it is also possible to adopt a penalized splines (P-spline) approach, i.e., use a large number of equi-spaced knots and introduce a penalty term to contain the model complexity and \say{mitigate} overfitting.  


\subsection{Exponentiated Polynomial Model}
\label{subsec:taylor}
In the exponentiated polynomial model we assume that the baseline hazard function is equal to
\begin{align*}
    \lambda_0(t | \bomega) = \exp\big\{\bomega^\top\mathbf{b}(t)\big\}
\end{align*}
where $\bomega$ is a $q$-dimensional vector of unknown  parameters and  $\mathbf{b}(t)=(b_0(t),\ldots,b_{q-1}(t))^\top$ with
\begin{align*}
    b_k(t) = t^k, \quad k = 0, \dots, q-1.
\end{align*}
So the function $\bomega^\top\mathbf{b}(t)$ is equal to a polynomial of the order $q-1$. The cumulative baseline hazard function can not be computed in a closed form: 
\begin{align*}
    \Lambda_0(t|\bomega) = \int_0^t \exp\big\{\bomega^\top\mathbf{b}(s)\big\} \ \rmd s 
\end{align*}
which must then be evaluated numerically (e.g., when maximizing the posterior density), e.g., via quadratures. Its gradient and the Hessian matrix equal 
\begin{align*}
     \nabla_{\bomega} \Lambda_0(t|\bomega) = \int_0^t \mathbf{b}(s) \exp\big\{\bomega^\top\mathbf{b}(s)\big\} \ \rmd s, 
\qquad \qquad     \nabla_{\bomega}^2 \Lambda_0(t|\bomega) = \int_0^t \mathbf{b}(s)\mathbf{b}(s)^\top \exp\big\{\bomega^\top\mathbf{b}(s)\big\} \ \rmd s 
\end{align*}
and $\nabla_{\bomega}^2 \log \{\lambda_0(t|\bomega)\} = \bm{0}$, the $q\times q$ matrix with zeroes.

In the BFI framework it is important that the models that are fitted in the local centers have the same polynomial order. Possibly, the researchers across the different local centers can agree upon the model, or the researcher in the central server can determine this. Since the models are nested, there is hardly any risk if the order is chosen slightly too high; the estimates of the parameters belonging to the redundant higher order terms will be close to zero. If the order is chosen too small, the estimated baseline hazard function will be less flexible than it should be. 

It might be difficult to choose an appropriate order beforehand. The order for the polynomial function can also be chosen based on statistical arguments. The idea is that in each local center the model with the \say{best} polynomial order is chosen based on a statistical procedure. For $q_\ell^\star$ the chosen order in center $\ell$, this yields the orders $q_1^\star, \ldots, q_L^\star$ in the $L$ centers. Since our initial assumption is that the models in all centers share the same baseline hazard function, the order of the model is set equal to the maximum of the local orders $q^{\star} = \max \{q^{\star}_1, \ldots, q^{\star}_L \}$. Next, in every center a model with order $q^\star$ is fitted and the $L$ estimated models are combined with the BFI methodology to obtain a single model for the merged data.
The motivation for taking the highest-order polynomial is that it should capture more complex features and details in the combined data and thus estimate the baseline hazard function more accurately. Moreover, the models are nested and the model with a higher order includes the models with lower orders. 

One way to select the optimal local order within a center is by using the likelihood ratio test. Let $\mathcal{M}_{q}$ be the model with a polynomial of order $q$. Note that the models form a sequence of nested models; model $\mathcal{M}_{0}$ is a special case of model $\mathcal{M}_{1}$, model $\mathcal{M}_{1}$ is embedded into $\mathcal{M}_{2}$, and so on. In every center, first the null hypothesis $H_0:\mathcal{M}_0=\mathcal{M}_1$ is tested against the alternative hypothesis $H_1:\mathcal{M}_0 \neq \mathcal{M}_1$. The model $\mathcal{M}_0$ is the simplest model where the baseline hazard function is constant over time (i.e., the exponential model), and model $\mathcal{M}_1$ assumes a baseline hazard function that is exponential in time. If $H_0$ is not rejected, then the procedure ends, and we choose model $\mathcal{M}_0$ as the \say{best} model given the data in the local center. Otherwise, the procedure continues with testing the null hypothesis $H_0:\mathcal{M}_1=\mathcal{M}_2$ against the alternative hypothesis $H_1:\mathcal{M}_1  \neq \mathcal{M}_2$, and so forth until the null hypothesis is not rejected.

The procedure described above would involve two rounds of communication between the central server and the local centers, since the order of each center must be communicated first to the central server in order to decide upon the final order. In order to avoid this and achieve a single round communication process, the local estimated model with the optimal order, as well as all fitted models with a higher polynomial order are sent to the central server. In the central server, the maximum order is computed and the local models with the corresponding orders are combined. 

\section{Simulation studies}
\label{sec: sim studies}

\subsection{Settings}
Suppose there are $L=3$ centers with data sets with sample sizes varying between 50, 100 and 500. For each patient the time-to-event is simulated from a multivariable Cox model with four covariates. The baseline hazard function correspond to the Weibull distribution with a scale parameter of $\exp(\omega_1)$ where $ \omega_1=-0.9$ and a shape parameter of $\exp(\omega_2)$ with $ \omega_2=1.8$. Within a patient (and across patients) the covariate values are independently and randomly generated from a standard normal distribution. The values of the regression coefficients are set equal to $\bbeta= (-0.6,-0.4,0.4,0.6)$ to achieve hazard ratios ranging from about 0.5 to 2. For generating the survival data with a predefined censoring rate of $30\%$, we used a small modified version of the methodology proposed by Wan\cite{Wan22} (see Appendix~\ref{appendix:Generation1}). 

For analysis, we consider six models: exponential (Exp), Weibull (Wei), Gompertz (Gom), exponentiated polynomial (Poly), piece-wise exponential with four intervals (PW4) and piece-wise exponential with eight intervals (PW8). Further, we assumed zero-mean multivariate Gaussian distribution as (relatively uninformative) priors with the inverse covariance matrix equal to a diagonal matrix $\mathbf{\Gamma}_{\bbeta}= \mathbf{\Gamma}_{\bbeta\ell}=10^{-2} \mathbf{I}_4$ and $\mathbf{\Gamma}_{\oomega}=\mathbf{\Gamma}_{\oomega\ell}=10^{-2} \mathbf{I}_q$ where $q$ refers to the number of model parameters for the baseline hazard function. For example, in the PW4 model $q=4$. For the Poly model we set a maximum value for $q$ to be $2$ (i.e., $\omega_0$ and $\omega_1$), corresponding to a first order polynomial. For each choice of the sample sizes in the centers, data sets are simulated and analyzed $B=100$ times.  

For comparison, we consider two more estimators which are commonly used in practice if the data from different centers can not be merged. The first estimator is referred to as the weighted average estimator (WAV). This estimator is found by combining the MAP estimators from the $L$ different data sets by taking a weighted average where the weights are based on the size of the data set. The estimator for the $k^{th}$ coordinate is defined as: 
\begin{align*}
\widehat{\beta}_{WAV,k}  &= \frac{1}{n} \sum_{\ell=1}^{L} n_\ell \widehat{\beta}_{\ell,k} ,
\end{align*}
where $n=\sum_{\ell=1}^{L} n_\ell$. The second extra estimator is simply the MAP estimator in the largest of the $L$ data sets. We refer to this estimator as $\widehat{\bbeta}_{Single}$ and to its $k^{th}$ coordinate as $\widehat{\beta}_{Single,k}$. This reflects the situation in which one performs the analysis in a single center only and one does not try to combine estimates or data.

\subsection{Measures to quantify performance}
\label{subsubsec: MSE}
With the BFI methodology we try to reconstruct from local inferences what we would have obtained if we had merged the data sets before doing the analysis. That means that the BFI estimators by definition cannot do better than the MAP estimators based on the combined data (the gold standard). Therefore, the parameter estimates obtained by the BFI approach are compared to those found after combining the data. Also the  estimates based on a single data set and the weighted average of local estimates are compared to the combined data estimate. 

Let, for a given model, $(\widehat{\bbeta}_{BFI}, \widehat{\oomega}_{BFI}, \widehat{\mathbf{M}}_{BFI})$ be the BFI-estimates and  $(\widehat{\bbeta}_{Com}, \widehat{\oomega}_{Com}, \widehat{\mathbf{M}}_{Com})$ the MAP estimates found after combining all data. To verify the performance of the BFI estimation method the mean squared error (MSE) is computed for each coefficient:
\begin{align*}
MSE_{\beta_k, BFI}  = \frac{1}{B} \sum_{b=1}^{B}  \big(\widehat{\beta}_{BFI, k}^{(b)} - \widehat{\beta}_{Com, k}^{(b)} \big)^2,
\end{align*}
where $\widehat{\beta}_{BFI, k}^{(b)}$ is the estimated value of the $k^{\text{th}}$ coefficient of $\bbeta$ using the BFI method in the $b^{\text{th}}$ iteration, and $\widehat{\beta}_{Com, k}^{(b)}$ is the estimated value of the $k^{\text{th}}$ coefficient using the combined data in the $b^{\text{th}}$ iteration. Similarly, we define the MSE for the two other estimators $\widehat{\beta}_{WAV,k}$ and $\widehat{\beta}_{Single,k}$ as follows:
\begin{align*}
MSE_{\beta_k, WAV}  = \frac{1}{B} \sum_{b=1}^{B}  \big(\widehat{\beta}_{WAV, k}^{(b)} - \widehat{\beta}_{Com, k}^{(b)} \big)^2, ~~~~~~~~~~~ MSE_{\beta_k, Single}  = \frac{1}{B} \sum_{b=1}^{B}  \big(\widehat{\beta}_{Single, k}^{(b)} - \widehat{\beta}_{Com, k}^{(b)} \big)^2.
\end{align*}

In all definitions of the MSE the estimate of interest is compared to the estimate that is based on the merged data. A small value means that there is hardly any loss when computing the corresponding estimator compared to what would have been obtained if all data had been merged. As a consequence of the above definitions, we do not yet measure model misspecification. A model that is misspecified can still have a small MSE, and the estimates from the true model may not necessarily have the smallest MSEs.    

The square root of a diagonal element of the inverse of $\widehat{\mathbf{M}}_{BFI}$ is an estimate of the standard deviation of the BFI estimator of the corresponding parameter. Its accuracy (compared to what would have been found if the data sets were merged) is estimated by the MSE:
\begin{align*}
MSE_{(\mathbf{M})_{kk},BFI} = \frac{1}{B}\sum_{b=1}^B  \bigg(\Big\{\big(\widehat{\mathbf{M}}_{BFI}^{(b)}\big)^{-1}\Big\}_{kk}^{\tfrac{1}{2}} - \Big\{\big(\widehat{\mathbf{M}}_{Com}^{(b)}\big)^{-1}\Big\}_{kk}^{\tfrac{1}{2}} ~\bigg)^2. 
\end{align*} 


Since the number of baseline hazard parameters $\oomega$  varies across the analysis models, the following MSE is employed to evaluate the efficacy of the BFI estimators of the baseline hazard functions:
\begin{align*}
MSE_{\Lambda_0,BFI}(t^*)  = \frac{1}{B} \sum_{b=1}^{B}  \big(\widehat{\Lambda}_{0,BFI}^{(b)}(t^*) - \widehat{\Lambda}_{0,Com}^{(b)}(t^*) \big)^2 .
\end{align*}
To quantify the difference between the curves $\widehat{\Lambda}_{0,BFI}(t^*)$ and $\widehat{\Lambda}_{0,Com}(t^*)$, we designate four distinct time points for $t^*$ such that the quantiles of the probability distribution of survival time of the distribution we simulate from corresponding to these points are $0.2$, $0.4$, $0.6$, and $0.8$. For the Weibull distribution with $\omega_1=-0.9$ and $\omega_2=1.8$, the time points are set to $t^*=(0.9056, 1.0385, 1.1438, 1.2554)$.

In the simulation study, the mean squared errors are computed for every analysis model, every model parameter,  for the three estimators and for multiple combinations of the sample sizes.

\subsection{Simulation results}
\label{subsec:results}
In this subsection, we present the results of our simulation study. First, for the different analysis models the MSEs of the regression parameters based on the three estimators have been computed. They are presented in the table \ref{tab:mseweib1}. Consistently, across all sample sizes, the models exhibited small MSEs for the regression coefficients computed with the BFI estimates, even for small sample sizes. This holds for all models, also if the model is misspecified. This indicates that there is hardly any loss when computing the estimates with the BFI methodology, compared to the estimates that would have been found after merging the data. Also the weighted average estimator shows low MSEs. The single center estimator seems to do worse for low sample sizes, but improves if the sample size in the center increases.  Further, as expected, for all models and all estimators, the MSEs tend to decrease for increasing sample sizes by decreasing variability of the estimates and possible decreasing overfitting bias. The MSEs for $\Lambda_0$ at four quantiles are shown in Table \ref{tab:mseweib2}. It can be seen that the curves estimated by the BFI methodology are close to those estimated after combining the data sets, particularly for large local samples. 

\begin{figure}
\centering
\includegraphics[scale=0.6]{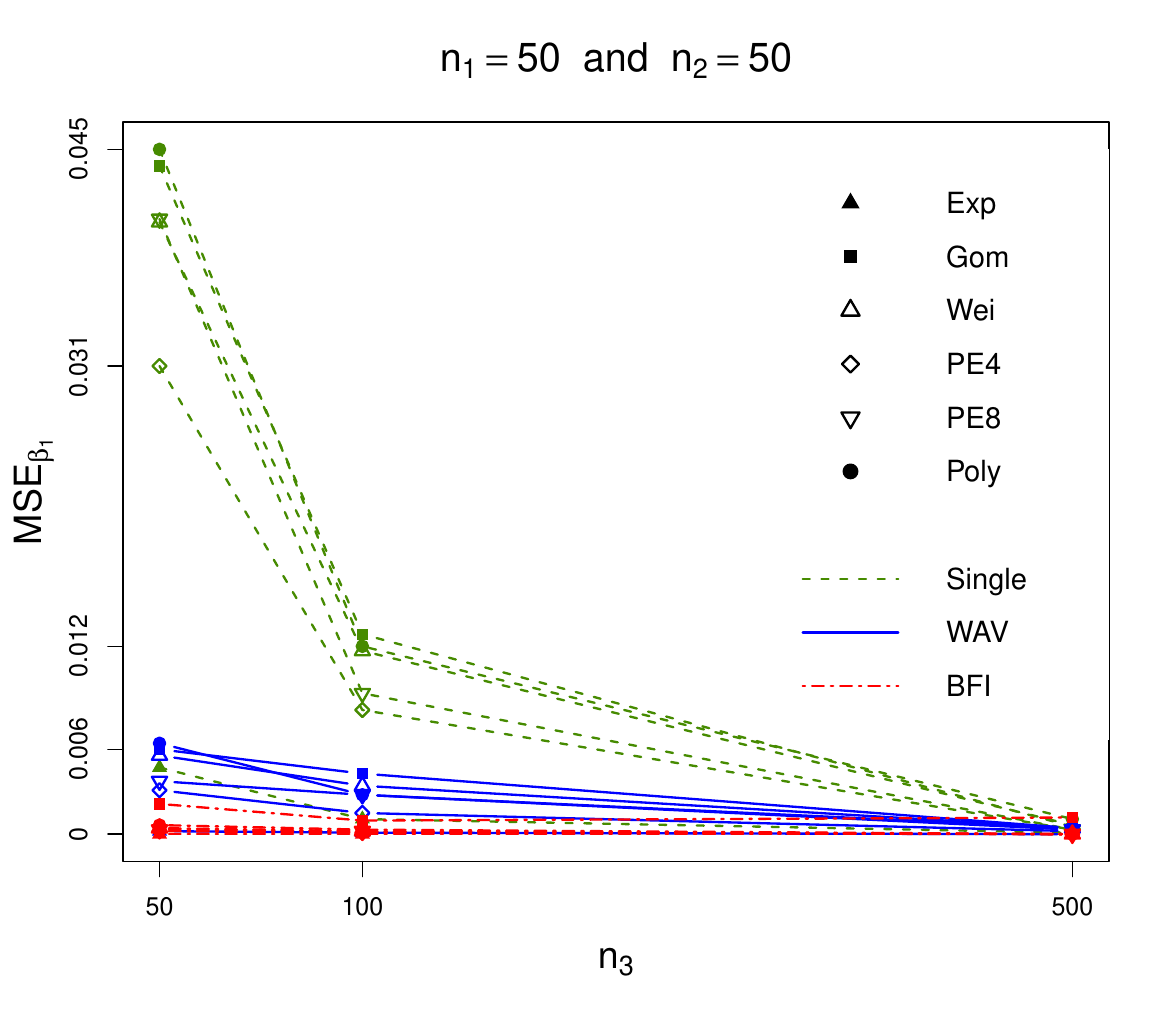}
\caption{The $MSE_{\beta_1, Single}$, $MSE_{\beta_1, WAV}$ and $MSE_{\beta_1, BFI}$ for different models and as a function of the sample size in the third center $n_3$. The sample size in center 1 and 2 ($n_1$ and $n_2$) are fixed to 50. Some curves overlap. The green dashed lines (`Single') correspond to the $MSE_{\beta_1, Single}$ computed based on the estimate in the center with the highest sample size. The blue solid lines (`WAV') correspond to the $MSE_{\beta_1, WAV}$ computed based on the weighted average of the estimates in the different centers. The red dashed lines (`BFI') correspond to the $MSE_{\beta_1, BFI}$ and is based on the BFI estimates. In all cases the MSE is computed for comparison with the estimate for $\beta_1$ computed based on the merged data. The survival data were generated from a Weibull distribution.
}
\label{fig:MSEbeta1EXP}
\end{figure}

To visualize some  of our results, we plotted the MSEs for the first regression parameter for different estimators and different models as a function of the sample size in the third center (Figure~\ref{fig:MSEbeta1EXP}). The sample sizes in the first and second center are fixed to 50. For small sample sizes in the three centers,  the gain of the BFI estimator compared to the singe center becomes clearly visible. For a large sample size in one of the centers  all estimators perform well; in those cases the largest data set (the third data set) simply dominates the full data set.

A second simulation study has been performed. In this study the life times were simulated from an exponential distribution (constant baseline hazard function). The results were very similar to the results as described above.

Increasing $\gamma$, the diagonal elements of the diagonal inverse covariance matrix of the prior, means stronger regularization and all parameter estimates are pushed towards zero. This leads to smaller MSE values. For $\gamma$ close to zero, the MAP estimator is approximately equal to the maximum likelihood estimator. A similar simulation study with $\gamma$ very small ($10^{-9}$) yielded a similar figure as Figure $\ref{fig:MSEbeta1EXP}$, but with slightly larger values of the MSEs.     

Better agreement between the estimates and those found after combining the data, does not necessarily mean that the estimated model is closer to the true model, as model misspecification is not measured. We therefore also computed the $MSE_{\bbeta,BFI}$ as defined before, but replaced the estimate that is based on the merged data in the definition, by the true values of the regression parameters in the true model. This yields the values $MSE_{\bbeta,BFI,True}$ given in Table \ref{tab:mseweib2}. Now, model misspecification is also measured. The MSE values for the Weibull model (the data were simulated from a Weibull model) are smaller than those for the other analysis models. Moreover, as expected, the MSEs are much larger now.

\section{Application: Salivary Gland Cancer}
\label{subsec:application}

\subsection{Data description}
Salivary gland cancer (SGC) is a rare and diverse malignancy with over 20 subtypes . The Radboud university medical center (Radboudumc) is a tertiary SGC expertise center in the Netherlands, receiving referrals from across the country. Data was collected in a real-world database and the used data set contained data from SGC patients referred to the Radboudumc from all 8 Dutch academic medical centers and tens of smaller non-academic hospitals, with 1 to 63 patients per center.\cite{Boon, Weijers} In general, the academic hospitals referred more patients than the smaller non-academic hospitals. Of 491 patients data was present. Of these patients, 205 have a date of death (an event). Our aim is to fit a multivariable model for the overall survival, defined as the time of diagnosis to death (of any cause), based on characteristics of the patient and the tumor. We will use the three estimators as defined before.

Based on literature and knowledge of our medical researchers, five covariates had been selected: age at diagnosis, sex, subtype, M-stage, and N-stage. The covariate ‘age’ is a continuous variable and is denoted in years, the covariate ‘sex’ is dichotomous. The variable ’subtype’ is categorical with three levels: Salivary duct carcinoma (reference level) and Adenoid cystic carcinoma, which are the most common subtypes, and the remainder, even more rare, subtypes combined into one category. The M-stage (metastasis stage) indicates whether the cancer has spread to other parts of the body. There are three categories: M0 (no distant metastasis, reference level), M1 (distant metastasis are present) and Mx (distant metastasis cannot be assessed). The covariate N-stage (regional lymph node stage) indicates whether the cancer is present in regional lymph nodes. In our dataset 5 categories were classified:  N0 (regional lymph nodes do not contain metastasis, reference group), N1 (there are cancer cells in 1, 2 or 3 nearby lymph nodes), N2 (there are cancer cells in 4 to 9 nearby lymph nodes), N3 (there are cancer cells in at least 10 nearby lymph nodes) and Nx (there is no information about the regional lymph nodes). Some of the categories have subcategories, but we ignored these in the analysis.\cite{Brierley} 

Categorical covariates can be included in the analysis by defining dummy variables. However, if the local sample size is small, it might happen that there are no patients in one or more categories of a categorical variable. The corresponding category can not be left out from the model, because this would lead to different models across the centers. Because we are working in a Bayesian setting, the MAP estimate of the regression parameter of the `empty category' can be estimated as the mean of the Gaussian prior (so zero). First and second derivatives of the log posterior density are also non-zero by the presence of the prior.

\subsection{Random allocation to medical centers}
In the first data analysis we randomly distribute all patients in the data set over three fictive medical centers with, respectively, 100, 150, and 241 patients of which less than 50\% had an event. The sample sizes, and thus the number of events, are relatively large in the three hospitals compared to the actual sample sizes in the centers. Still, compared to the number of model parameters, the number of events are low.
In every of these three data sets (i.e., medical centers), we compute the MAP estimates of the parameters in the models described in Section 3 (Weibull, Gompertz, piecewise constant, exponentiated polynomial). For the parameter prior we took a multivariate Gaussian distribution with mean zero and a diagonal inverse covariance matrix with $\gamma=0.01$ on the diagonal. A value $\gamma=0.01$ corresponds to a variance of 100. So the MAP estimates are close to the maximum likelihood estimates. Next, for every model the MAP estimates from the three centers are combined with the BFI methodology (with the same prior for the combined data set) and the weighted average method. The single center estimator uses the data from the center with 241 patients. For all models and all patients the linear combinations $\bz_{\ell i}^\top \widehat{\bbeta}_{BFI}, \bz_{\ell i}^\top \widehat{\bbeta}_{WAV}$ and $\bz_{\ell i}^\top\widehat{\bbeta}_{Single}$ are computed. Also, for every model, the MAP estimates of the model parameters were determined based on the merged data set and for every patient the corresponding linear combination $\bz_{\ell i}^\top \widehat{\bbeta}_{Com}$ was computed. For every model, scatter plots of the points $(\bz_{\ell i}^\top \widehat{\bbeta}_{BFI},\bz_{\ell i}^\top\widehat{\bbeta}_{Com})$, of the points $(\bz_{\ell i}^\top \widehat{\bbeta}_{WAV},\bz_{\ell i}^\top\widehat{\bbeta}_{Com})$ and of $(\bz_{\ell i}^\top \widehat{\bbeta}_{Single},\bz_{\ell i}^\top\widehat{\bbeta}_{Com})$ were made. Since we aim to compute from the separate inference results in the local centers what would have been obtained if the data sets had been merged before the analysis, finding all points on the line $y=x$ will correspond to a perfect fit. For the Weibull model, the three scatter plots and the estimates of $\Lambda_0$ are given in Figure \ref{fig:3hospital_wei}. The plots for the other models are given in Figure \ref{fig:3hospital} in the appendix.

From the scatter plots, we see that for all models the BFI methodology performs very well; the BFI estimates $\bz_{\ell i}^\top \widehat{\bbeta}_{BFI}$ are almost exactly equal to the corresponding estimates based on the merged data. There is a minor rotation visible in the plot, possibly due to overfitting in the centers. The weighted average method also performs well for all models, but the accuracy is clearly lower than that of BFI. The performance of the single center estimator is not that good: the accuracy is low and the cloud with points seems to be rotated, possibly due to overfitting. 

The plots of the estimates of $\Lambda_0$ for the Weibull model in Figure \ref{fig:3hospital_wei} and for the other models in Figure \ref{fig:3hospital} show that the BFI estimator performs well; the estimated curve is very close to the estimated curve based on the combined data. The weighted average estimator performs reasonably well, but the single center estimator is quite far from the estimated curve based on all data.  

We repeated the whole procedure several times and the conclusions were found to be similar. 

\begin{figure}
\centering
\includegraphics[scale=0.5]{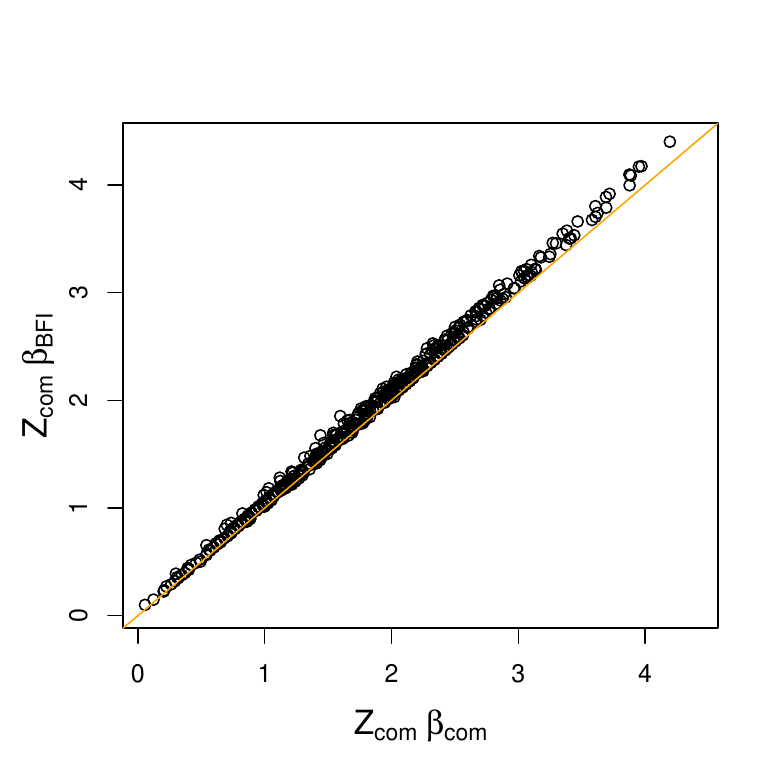}
\includegraphics[scale=0.5]{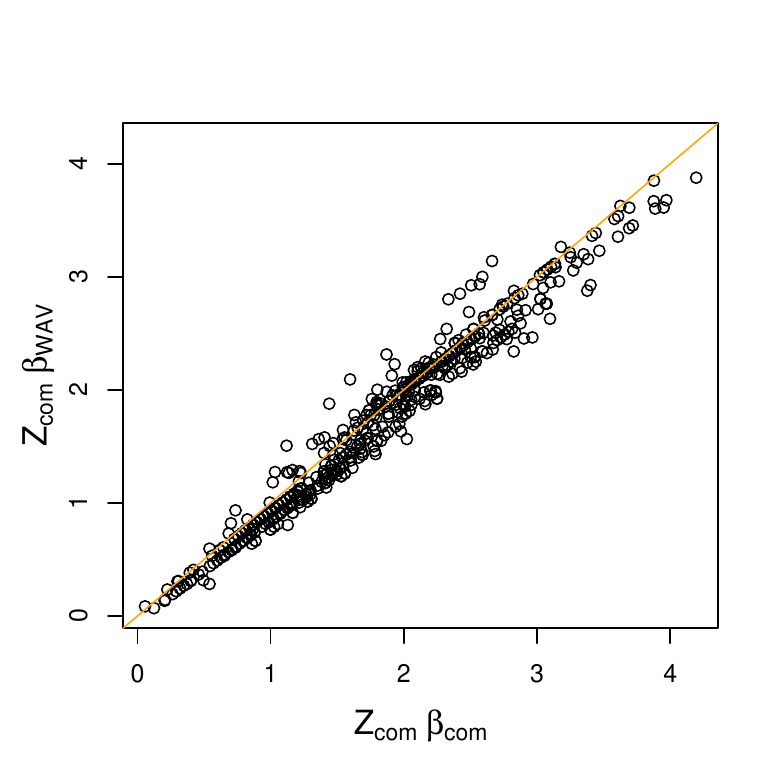}\\[-8pt]
\includegraphics[scale=0.5]{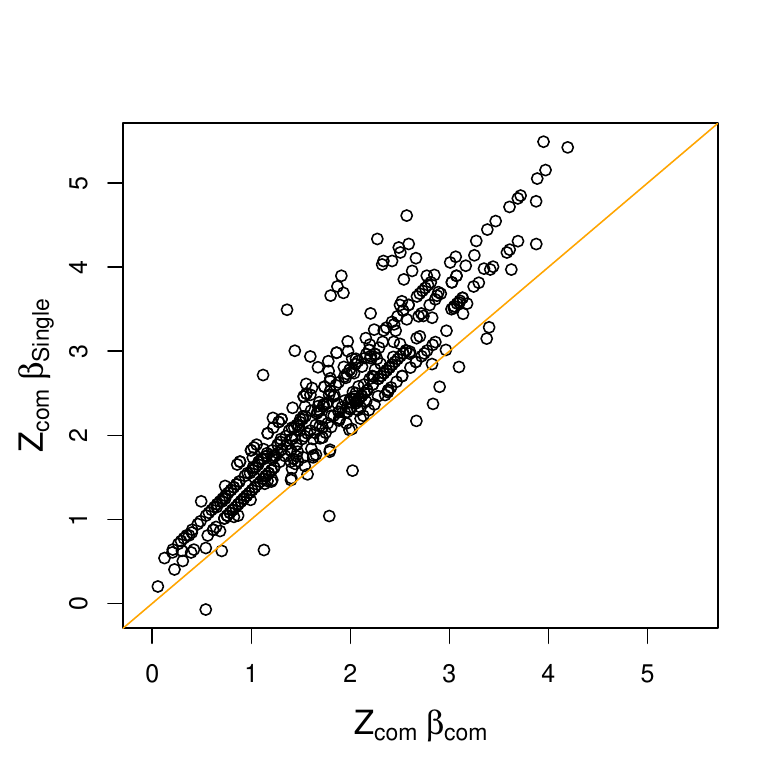}
\includegraphics[scale=0.5]{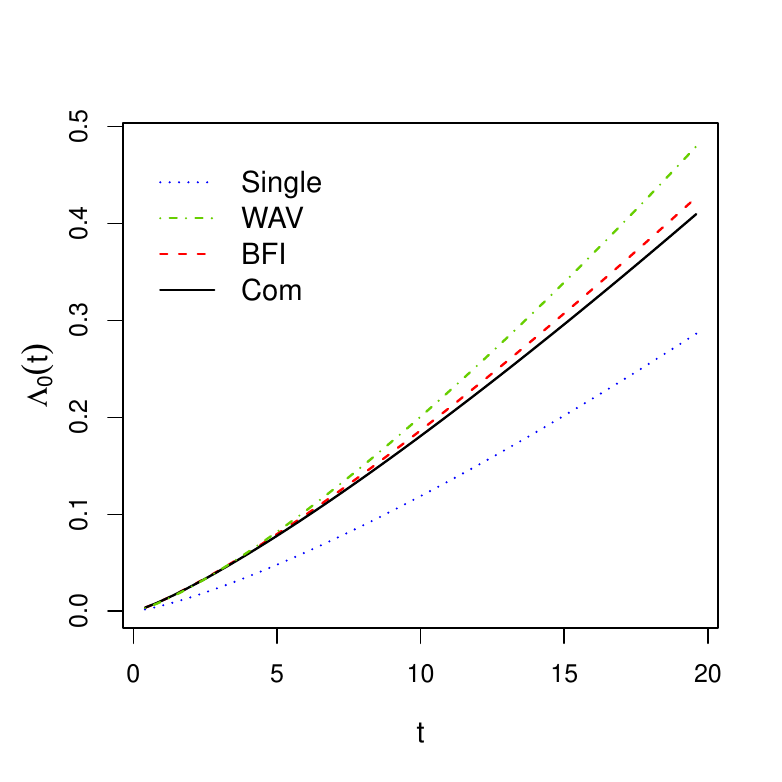}
\caption{All patients have been randomly allocated to three hospitals. Scatter plots of $\bz^\top \widehat{\bbeta}_{BFI}$, $\bz^\top \widehat{\bbeta}_{WAV}$ and $\bz^\top \widehat{\bbeta}_{Single}$ against $\bz^\top \widehat{\bbeta}_{Com}$, respectively, for the Weibull model. Fourth plot: estimates of $\Lambda_0$ in the Weibull model. The priors are zero mean Gaussian distributions with diagonal inverse covariance matrices with $\gamma=0.01$ on the diagonal.}
\label{fig:3hospital_wei}
\end{figure}

\subsection{Four hospitals}
\begin{figure}
\centering
\includegraphics[scale=0.5]{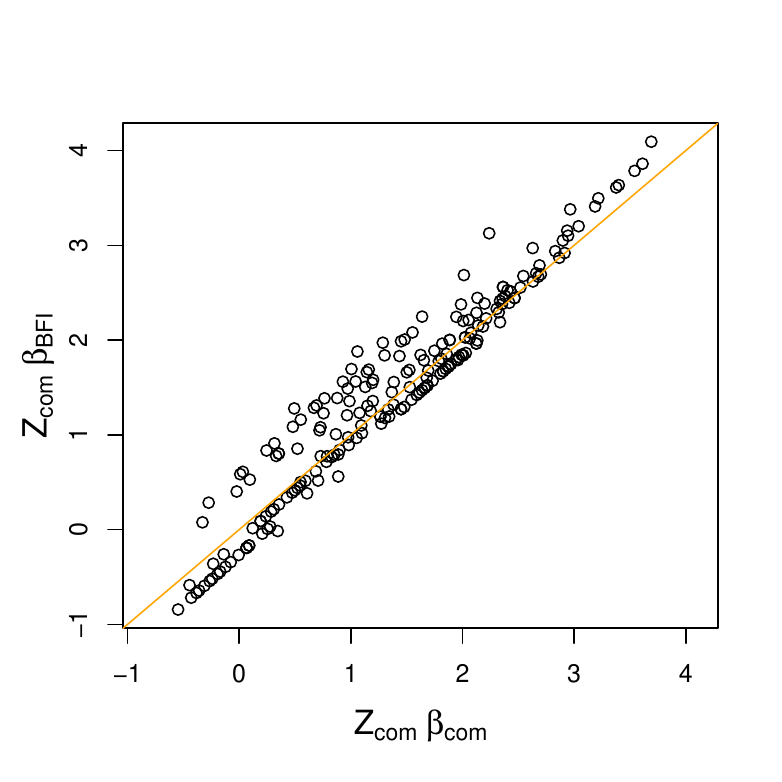}
\includegraphics[scale=0.5]{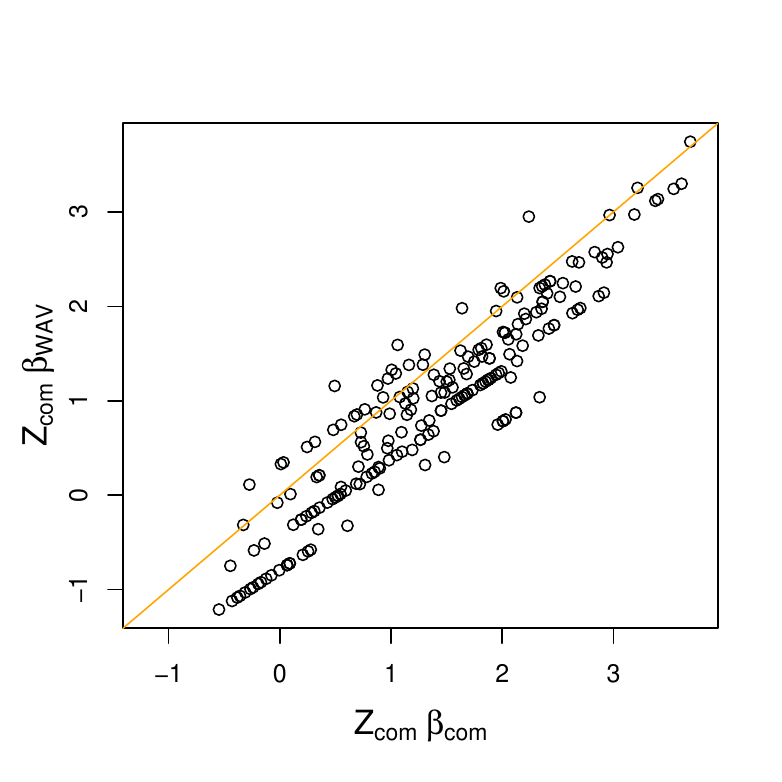}\\
\includegraphics[scale=0.5]{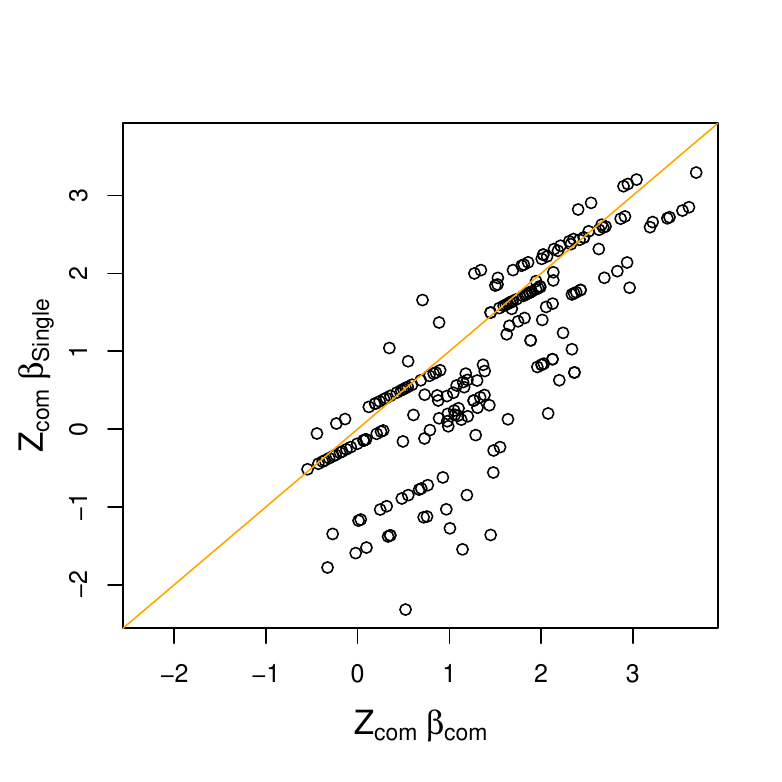}
\includegraphics[scale=0.5]{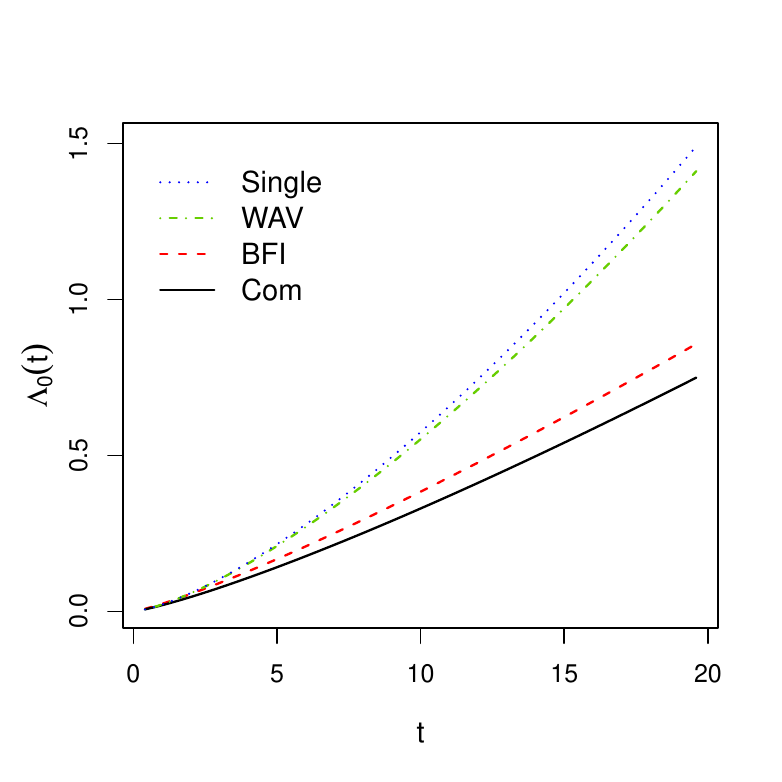}
\caption{Data from four hospitals have been used. Scatter plots of $\bz^\top \widehat{\bbeta}_{BFI}$, $\bz^\top \widehat{\bbeta}_{WAV}$ and $\bz^\top \widehat{\bbeta}_{Single}$ against $\bz^\top \widehat{\bbeta}_{Com}$, respectively, for the Weibull model. Fourth plot: estimates of $\Lambda_0$ in the Weibull model. The priors equal zero mean Gaussian distributions with diagonal inverse covariance matrices with $\gamma=0.1$ on the diagonal. }
\label{fig:4hospital_wei}
\end{figure}

We selected all centers which referred more than 30 patients each. Four centers satisfied this condition. Their sample sizes  are 42, 43, 60 and 63, with, respectively, 18, 12, 39 and 28 events. We performed the BFI protocol with the Weibull, Gompertz, piecewise constant (with four intervals) and the exponentiated polynomial model (with threshold 0.10). For the prior we took a multivariate Gaussian distribution with mean zero and a diagonal inverse covariance matrix with $\gamma=0.1$ on the diagonal. A value $\gamma=0.1$ corresponds to a variance of 10. Since the number of events are really small compared to the number of model parameters, we used a larger value of $\gamma$ to overcome overfitting. The same scatter plots as in the previous subsection were made (see Figure \ref{fig:4hospital_wei} for the Weibull model and Figure \ref{fig:4hospital} for the other models). From the figures we see that the BFI estimates perform reasonably well for all models. The points are nicely scattered around the $y=x$ line. Although for most of our  models the cumulative baseline hazard function seems to be overestimated, this effect is only minor. The weighted average estimator performs less well. It seems that the linear predictor $\bz^T \widehat{\bbeta}_{WAV}$ is underestimated; most of the points are located below the $y=x$ line. However, in the plots with the estimates of the cumulative baseline hazard function we see that the estimate for the weighted average strategy clearly overestimates the cumulative baseline hazard function. Possibly the biases in the two estimators compensate each other. However, as our aim is to approximate the estimates that would have been found if the data had been merged, we can conclude that the weighted average estimator didn't perform well in this data example. The single center estimator is based on the data from the largest data set, so based on data of 63 patients only. Again, we compare the estimates with those that were obtained after merging the data from the four medical centers. From the plots it is clearly visible that the estimator has a poor performance. 

In the two smallest datasets, consisting of data from 42 and 43 patients, some variable categories had no patients. The corresponding regression parameters were estimated as zero as this is the mean of the prior distribution.   

\section{Discussion}
\label{sec:discussion}
In this paper, the BFI methodology for GLMs is extended to survival models. Survival models are more complex than GLMs by the presence of the unknown baseline hazard function. Multiple parametric hazard functions of different complexity have been considered. Leaving the shape of the baseline completely unconstrained seems to be impossible without sharing patients’ survival data with the central server. We considered three baseline hazard functions of fixed shape: the exponential, the Weibull and the Gompertz distribution. A disadvantage of these models is their low flexibility. However, especially if the local data sets are small, such assumptions are necessary to be able to fit the model. The exponentiated polynomial approach gives more flexibility. The order of the polynomial function can be fixed beforehand, but can also be determined based on the data via a testing procedure. The piecewise constant baseline hazard function is also flexible if the number of intervals is not too small. We considered the situation with four and eight intervals. The position of the knots where chosen beforehand. Usually this is not a problem as the researchers in the local centers are allowed to have contact with each other to discuss model assumptions like the location of the knots. Which model is best in practice depends on the situation. If local sample sizes are large and the shape of the baseline hazard function is completely unknown, the polynomial and piecewise constant hazard functions would be preferred above the lower dimensional alternatives. However, if the local samples sizes are low, one should be careful using the piecewise exponential hazard function with many intervals. 

In the simulation study, the performance of the BFI estimators is compared to that of alternative estimators that could be considered for combining the local parameter estimators: namely the weighted average of the estimates, and using the estimates from the single center with the largest number of data. The results show that the BFI estimator outperforms the other estimators, but the differences in performance decrease if the local sample sizes increase (as one would anticipate). A better performance of the BFI estimator compared to the estimator based on the data of single center was expected, as the BFI estimator uses more data, and thus more information to estimate the parameters. The BFI estimators of the parameters are actually determined by a linear combination of the local estimators, but the weights are different from the weighted average estimator. Apparently this BFI combination leads to better estimators. The BFI estimator aims to approach the estimates that would have been obtained in the merged data set. That means that possible overfitting errors in the local centers (due to small sample sizes) are (partly) filtered out.   

In practical applications one needs to choose the inverse covariance matrix of the prior distribution. If the matrix equals a diagonal matrix with the value $\gamma$ on the diagonal, this $\gamma$ coincides with the penalization parameter in a ridge regression. However, in the BFI methodology a prior is chosen for every model parameter, whereas with ridge regression only the regression parameters are penalized. The higher the value of $\gamma$, the stronger the penalization will be. If only one data set is available, the best value of $\gamma$ could be selected via cross validation or other methods. In the BFI setting with multiple local centers, every center could choose its own value. The BFI strategy is constructed in such a way that the BFI estimates are approximately equal to the estimates obtained from the analysis based on the combined data set, where the sample size is higher (by definition) and less overfitting is expected to be present in the estimates. Therefore, it is not clear beforehand whether there is merit  in determining the best values of $\gamma$ in the local centers and, thus, use different values across the centers. This will be the topic of our next project.   

In the present paper it is assumed that the patients’ populations do not differ significantly across the centers. This is a strong assumption, since centers may be located in different areas and even different countries. For instance, one can imagine that the age distribution of patients in hospitals located in big cities is different from the one in hospitals in more rural areas, or that patients in academic hospitals are more serious ill than those in peripheral hospitals. This heterogeneity across the populations leads to different values of the parameters of the distributions of the covariates. Under the assumption that the regression parameters and the parameters that determine the baseline hazard function are statistically independent of the parameters for the covariate distribution, this type of heterogeneity does not affect the estimation of the regression parameters and the parameters for the baseline hazard function. 

We have also assumed that the baseline hazard functions are identical across centers. However, stratified BFI models can also be constructed, but sufficient data in the centers should be available. However, one should keep in mind that the BFI methodology was proposed to enlarge our statistical power for estimating the model parameters by increasing the ratio of sample size versus number of parameters. If there is expected to be strong hetereogeneity across locations it might be better to fit a BFI model only for centers for which the models are expected to be reasonably similar.

When applying the BFI methodology to data in different centers, it is important that the local models include the same covariates. Otherwise, combining the estimated local models is not sensible. This makes covariate selection difficult. An option could be to consider a multi-cycle procedure that visit the local centers multiple times to determine the best selection of the covariates based on a pre-defined criterion. An alternative would be to estimate regression models with every possible subset of covariates in every local center, send all estimated models to the central server and perform a model selection strategy at the central server. These approaches will be considered in more detail in one of our next projects.    

In the data example we used a data set of salivary gland cancer patients who were referred to the Radboud university medical center (Nijmegen, the Netherlands) for treatment. We used the hospital were the diagnoses had taken place to define the different medical centers. Since all data are available it was possible to measure and compare the performance of the different estimators. In reality this is not possible as the data can not be merged. With the results from this paper and from the previously published papers, we are convinced that the methodology works well for GLMs and time-to-event models.\cite{Jonkera,Jonkerb} Currently a project has been started to apply the methodology for estimation with data sets that are not combined. With the BFI methodology collaboration between researchers becomes easier and research proceeds faster, as no time-consuming data transfer agreements are necessary.

\bigskip

\noindent
{\bf{Acknowledgements}}\\
We thank Prof. dr. Carla van Herpen and Niels van Ruitenbeek for making the data on salivary cancer available. 

\bigskip

\noindent
{\bf{Software}}\\
The R package ``BFI'' has been developed for applying the BFI methodology for generalized linear models and the survival models described in this paper. The package and a tutorial are available at github: {\texttt {https://hassanpazira.github.io/BFI}}.

\bigskip

\noindent
{\bf{Data availability statement}}\\
The patient data can not be shared, as the data use is restricted due to ethical requirements.

\bigskip

\noindent
{\bf{Funding}}\\
This research was supported by an unrestricted grant of Stichting Hanarth Fonds, The Netherlands.

\appendix
\section{: Generating survival data with predefined censoring rates}
\label{appendix:Generation1}
Here, we explain the methodology for simulating right-censored survival data from a Weibull survival distribution, censoring times from a uniform distribution, and covariate values from a Gaussian distribution. Before simulating data, we need to determine a parameter that controls the proportion of censored subjects in the final data (called the censoring rate). This is done in three steps \cite{Wan22}: 1) Determine the censoring probability for each subject based on subject-specific covariates; 2) Derive a censoring rate function for the study population by marginalizing covariates from the patient censoring probability function; 3) Solve the censoring parameter within the censoring rate function to achieve a specified censoring proportion. Once we have the censoring parameter, we can proceed with simulating the data. 

We assume that the censoring time has density function $C \sim g(c|u_1,u_2)=\textit{Uniform}(u_1, u_2)$ where $u_1$ is known and $u_2>u_1$ is an unknown parameter, and that the event time $Y$ has the Weibull distribution. We let $T=\min(Y,C)$ be the observed follow-up time, $C$ be independent of $Y$, and $\Delta=I(Y\leq C)$ be the censoring indicator.
The patient specific covariates $\bz_i$ are  generated from the multivariate Gaussian distribution $\bz_i \sim \mathcal{N} (\textbf{0}_p, \sigma^2 \, \mathbf{I}_p)$. 

By assuming the baseline hazard from $\textit{Weibull}(\omega_2,1/\omega_1)$, with $ \lambda_0(t|, \omega_1,\omega_2)=\omega_1 \omega_2 t^{\omega_2-1}$, the patient censoring probability for the subject $i$ can be expressed as
\begin{align*}
\mathbb{P}(\Delta=0|\psi_i, u_1, u_2,\omega_1,\omega_2) &= \mathbb{P}(u_1 \leq C \leq u_2, C \leq Y \leq \infty) = \int_{0}^{\infty} g(c|u_1,u_2) \int_{c}^{\infty} f_{Y}(t|\omega_1,\omega_2)~ \rmd t \rmd c \\
&=\frac{1}{u_2-u_1}\int_{u_1}^{u_2}  \exp(- \psi_i \, \Lambda_0(c| \omega_1,\omega_2)/\omega_1 ) \rmd c = \frac{1}{u_2-u_1} \int_{u_1}^{u_2}  \exp(- \psi_i c^{\omega_2} ) \rmd c \\
&=\frac{\psi_i^{-1/\omega_2}}{(u_2-u_1)\omega_2} \int_{\psi_{i} u_1^{\omega_2}}^{\psi_i u_2^{\omega_2}} s^{1/b-1} \exp(- s ) \rmd s, \quad s=\psi_i c^{\omega_2} , \\
&=\frac{\psi_i^{-1/\omega_2}}{(u_2-u_1)\omega_2} \Big[ \Gamma(1/b , \psi_i u_2^{\omega_2}) - \Gamma(1/b,\psi_{i} u_1^{\omega_2}) \Big], \\
\end{align*}
where $\psi_i =\omega_1 \exp(\bz_i^\top\bbeta) \sim lnN(\ln{\omega_1},\sigma^2\sum_{j=1}^p \beta_j^2)$ has a lognormal distribution, and $\Gamma(\cdot,\cdot)$ is a lower incomplete gamma function. 
The censoring parameter $u_2$ needs to be determined to yield the desirable censoring rates in simulated survival data ($\pi$) by numerically solving the equations below with respect to $u_2$:
\begin{eqnarray*}
0=\mathbb{P}(\Delta=0|u_1, u_2,\omega_1,\omega_2) - \pi&=& \mathbb{E}_{\psi_i}(\mathbb{P}(\Delta=0|\psi_i, u_1, u_2,\omega_1,\omega_2))  - \pi \\
&=&\int_{0}^{\infty} \mathbb{P}(\Delta=0|s, u_1, u_2,\omega_1,\omega_2) f_{\psi_i}(s) ds  - \pi.
\end{eqnarray*}
By having the value of $u_2$, the survival data set can be simulated with predefined censoring rates $\pi$.

\section{: Tables and Figures describing simulation studies and data analysis}
\label{app: MSE sim studies}

\begin{figure}[h]
\centering
\includegraphics[scale=0.4]{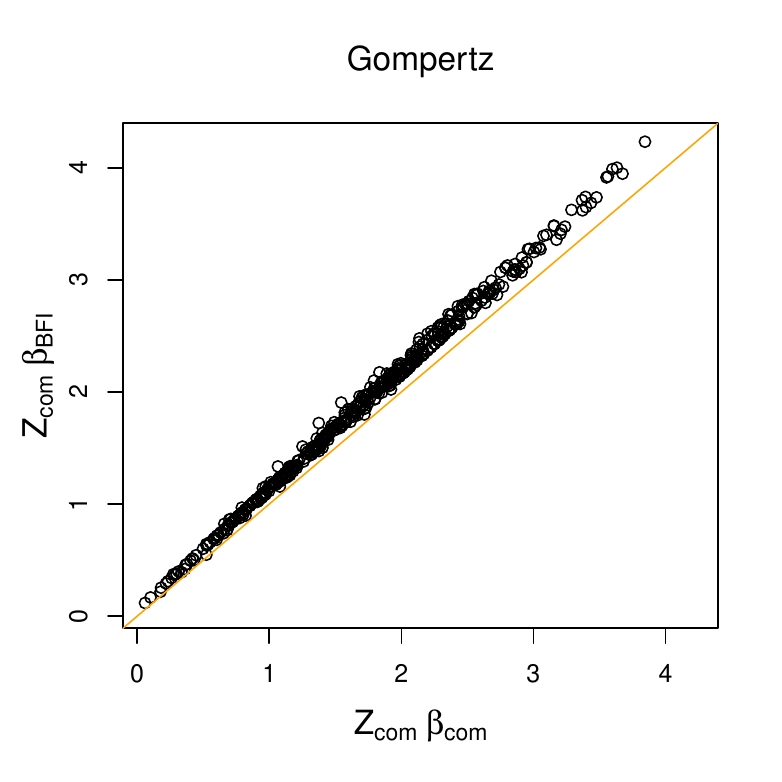}
\includegraphics[scale=0.4]{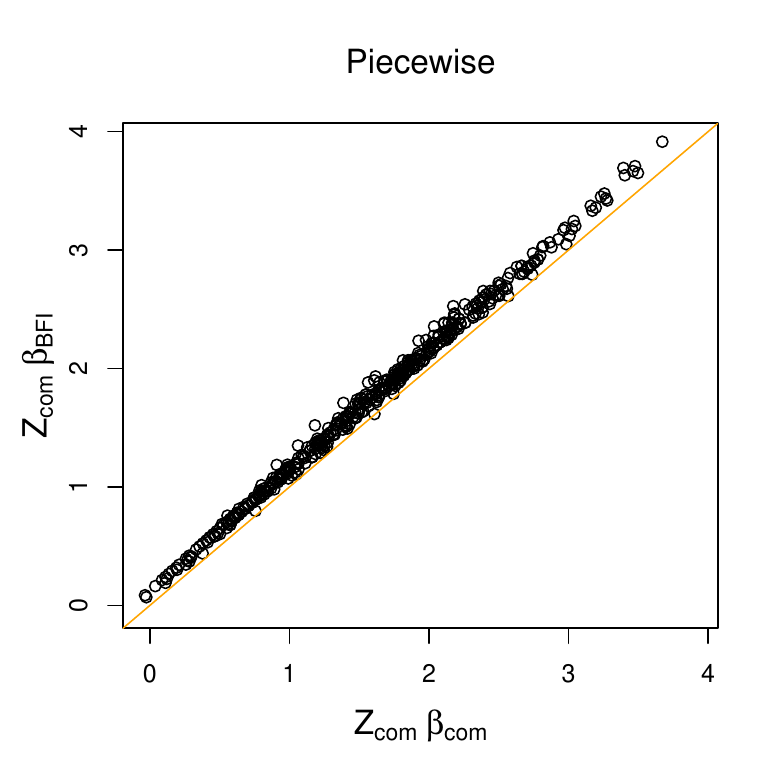}
\includegraphics[scale=0.4]{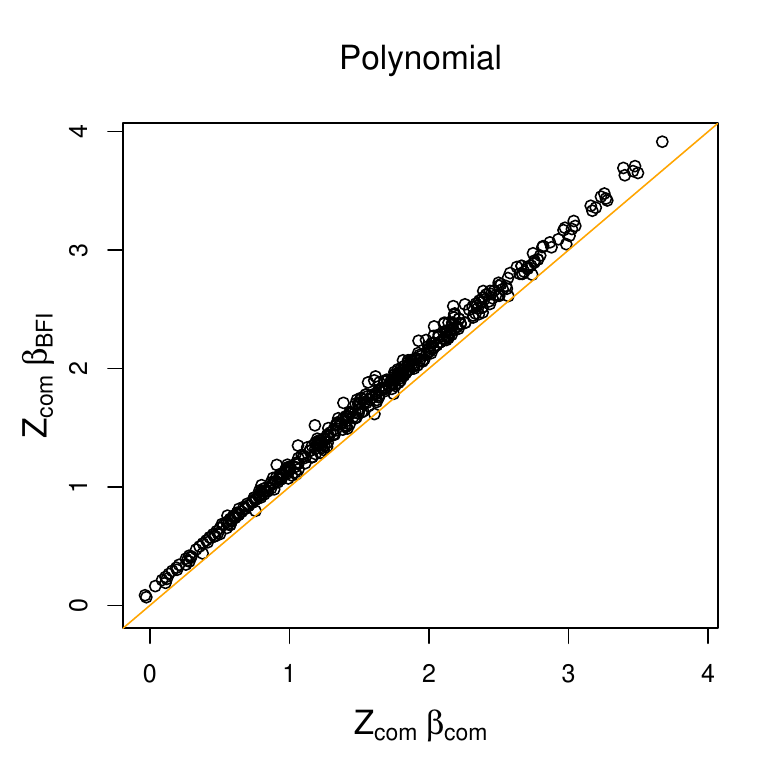}
\\[-8pt]
\includegraphics[scale=0.4]{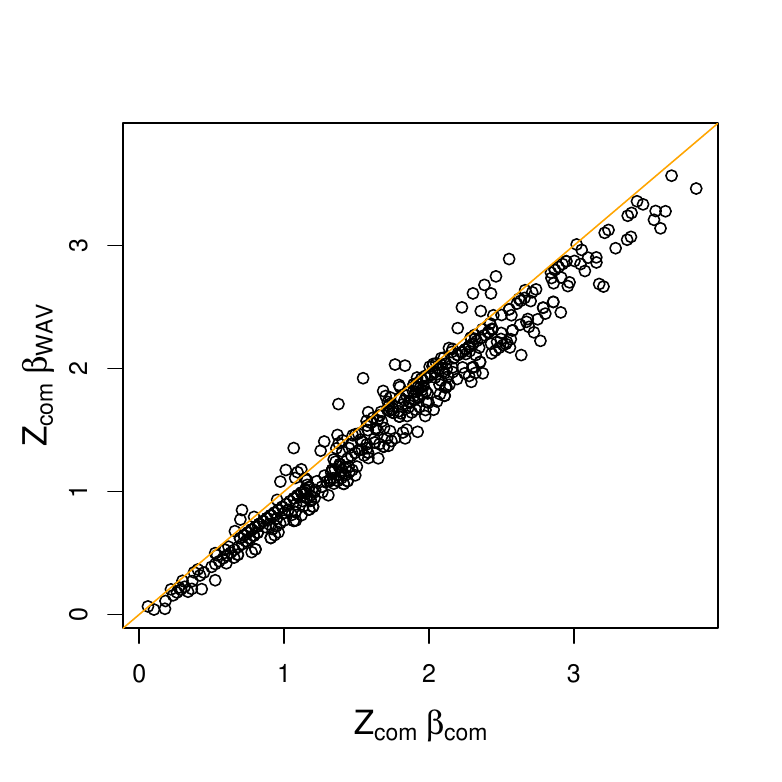} 
\includegraphics[scale=0.4]{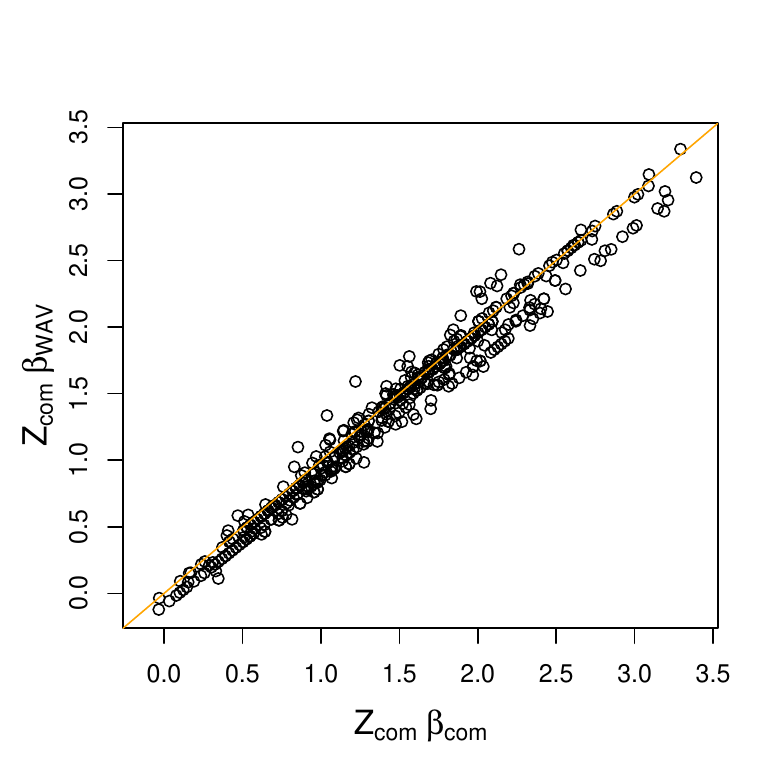} 
\includegraphics[scale=0.4]{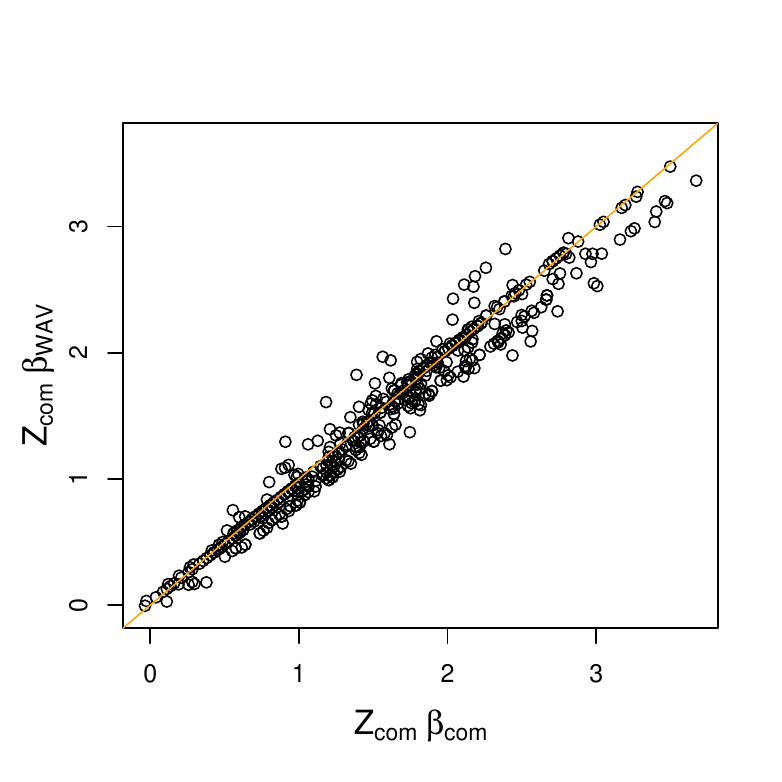} 
\\[-8pt]
\includegraphics[scale=0.4]{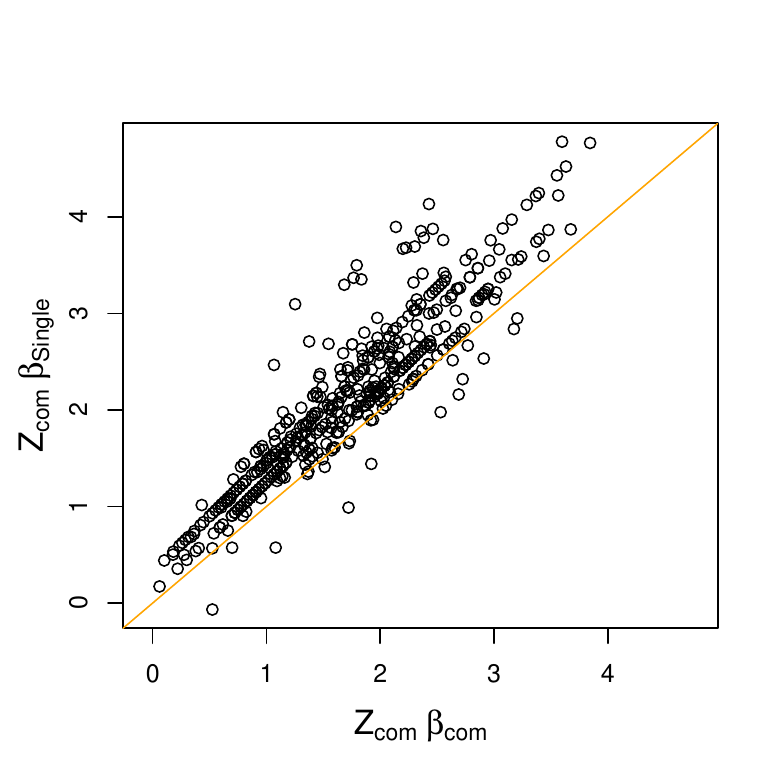}
\includegraphics[scale=0.4]{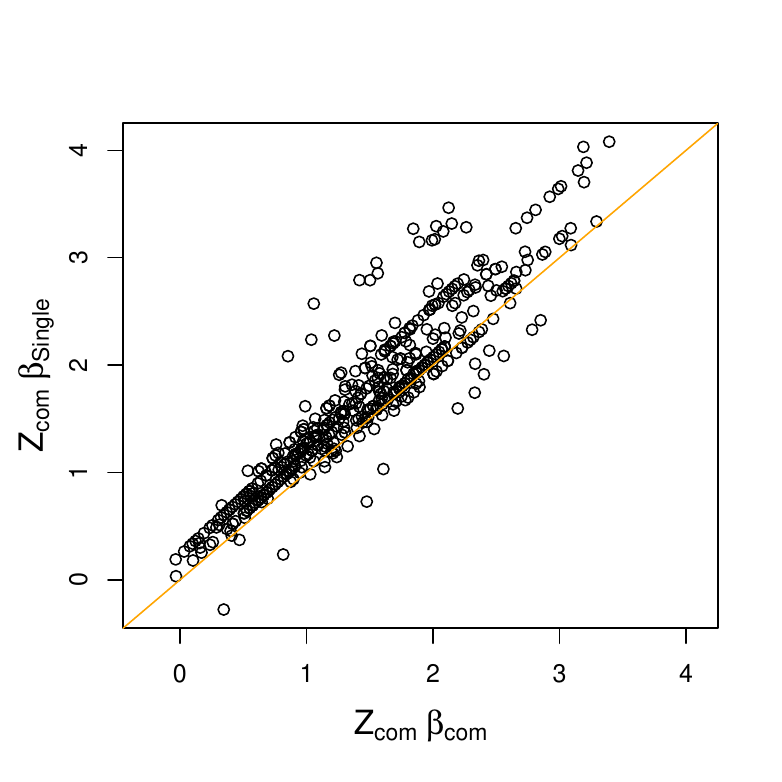}
\includegraphics[scale=0.4]{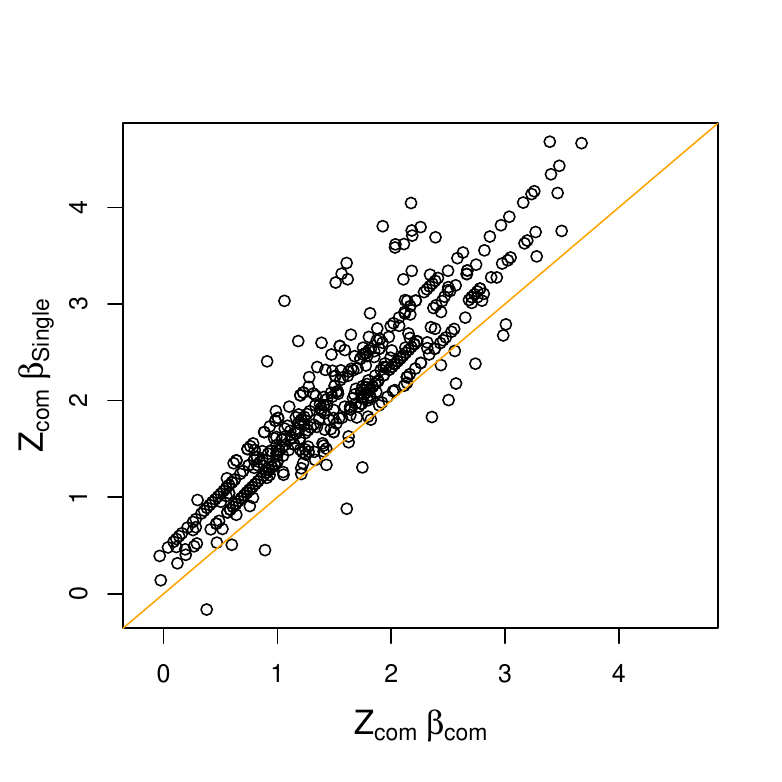}
\\[-8pt]
\includegraphics[scale=0.4]{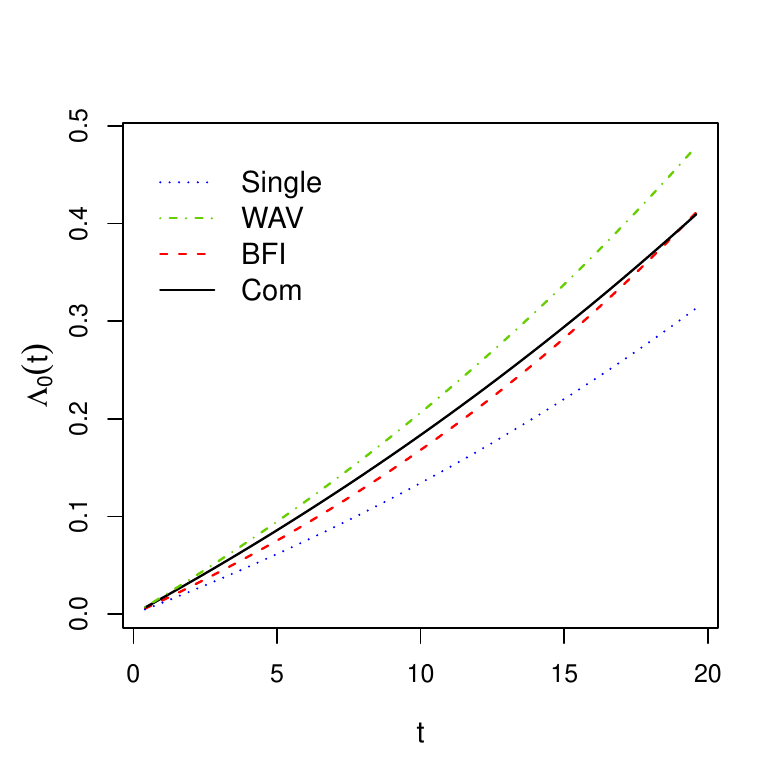}
\includegraphics[scale=0.4]{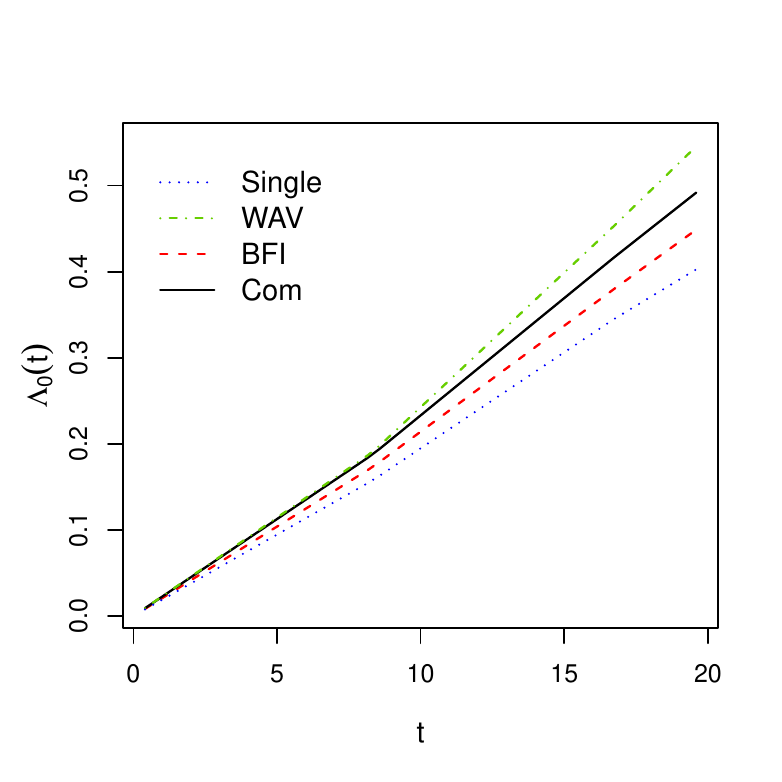}
\includegraphics[scale=0.4]{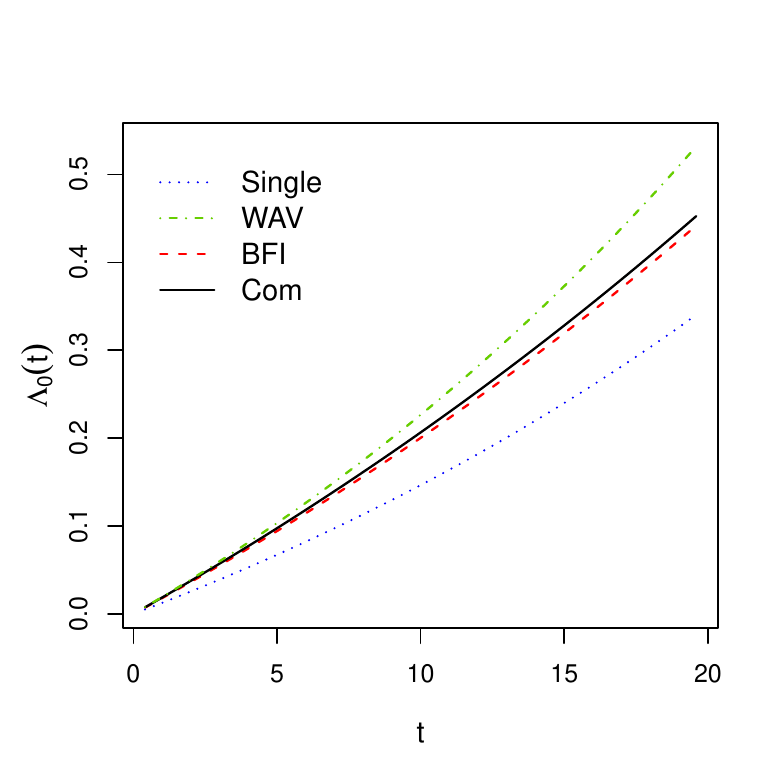}
\caption{All patients were randomly allocated to three medical centers. Scatter plots of $\bz^\top \widehat{\bbeta}_{BFI}$, $\bz^\top \widehat{\bbeta}_{WAV}$ and $\bz^\top \widehat{\bbeta}_{Single}$ against $\bz^\top \widehat{\bbeta}_{Com}$ in the first, second and third row, respectively, for three models: Gompertz (first column), piecewise constant with four intervals (second column), exponentiated polynomial (third column). Fourth row: estimates of $\Lambda_0$ in the three models: Gompertz (first plot), piecewise constant (second plot), exponentiated polynomial (third plot). The priors equal zero mean Gaussian distributions with diagonal inverse covariance matrices with $\gamma=0.01$ on the diagonal.}
\label{fig:3hospital}
\end{figure}

\begin{figure}[h]
\centering
\includegraphics[scale=0.4]{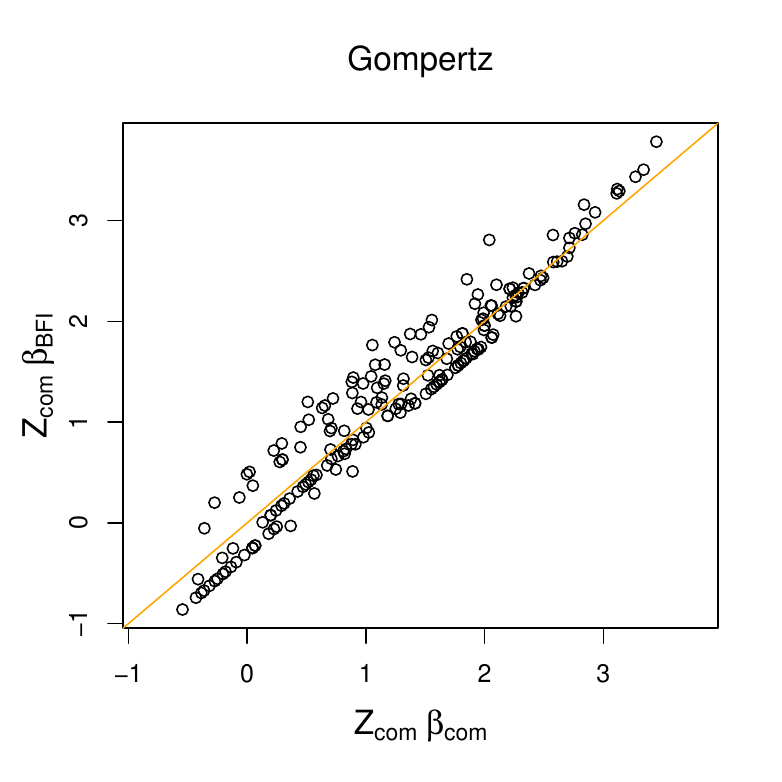}
\includegraphics[scale=0.4]{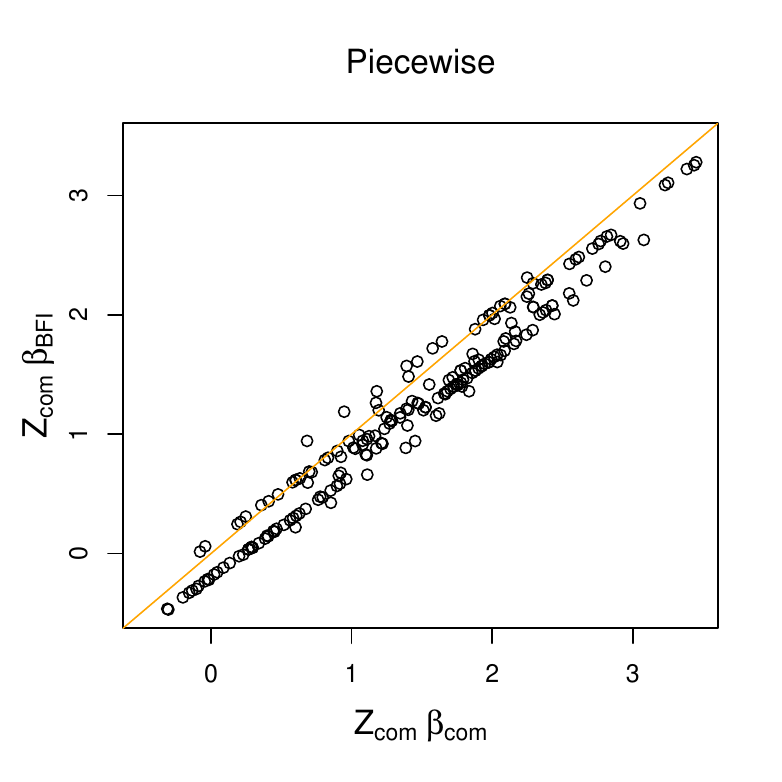}
\includegraphics[scale=0.4]{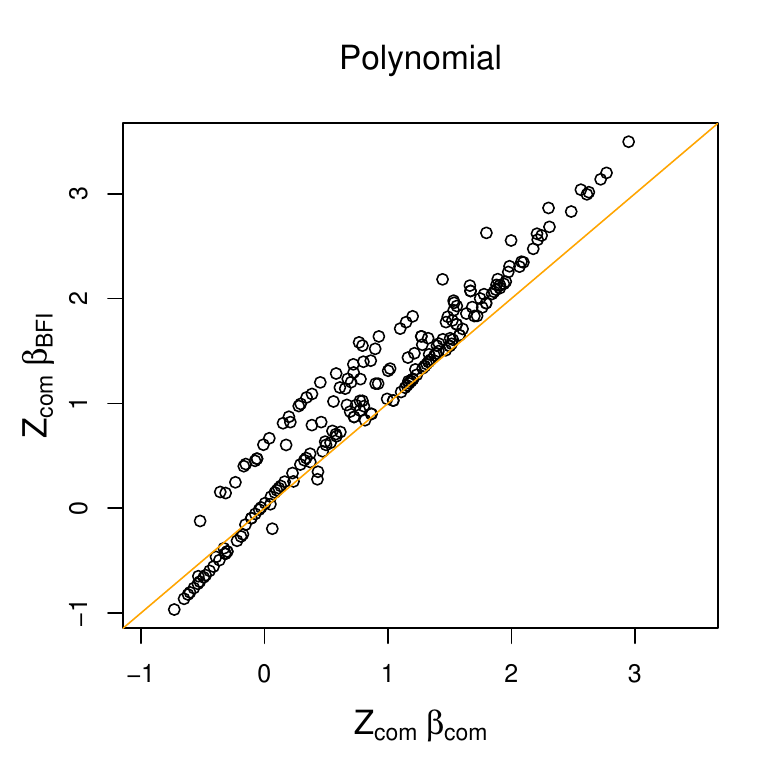}
\\[-8pt]
\includegraphics[scale=0.4]{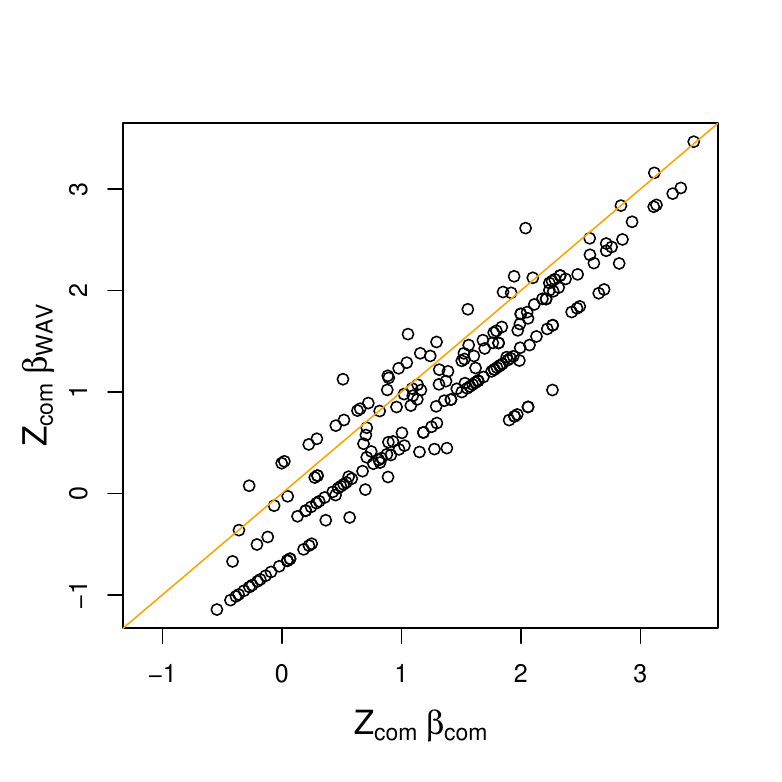}
\includegraphics[scale=0.4]{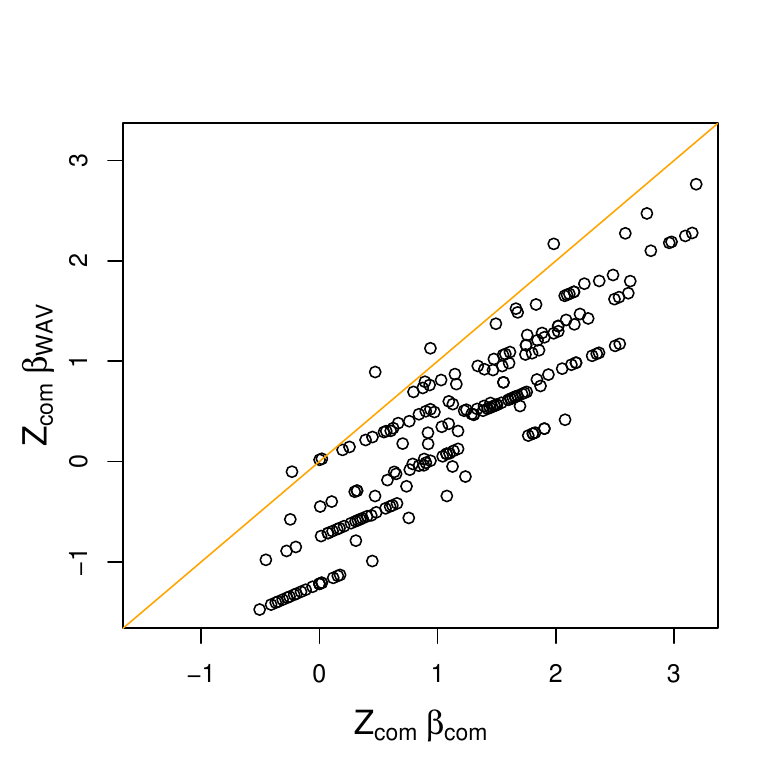}
\includegraphics[scale=0.4]{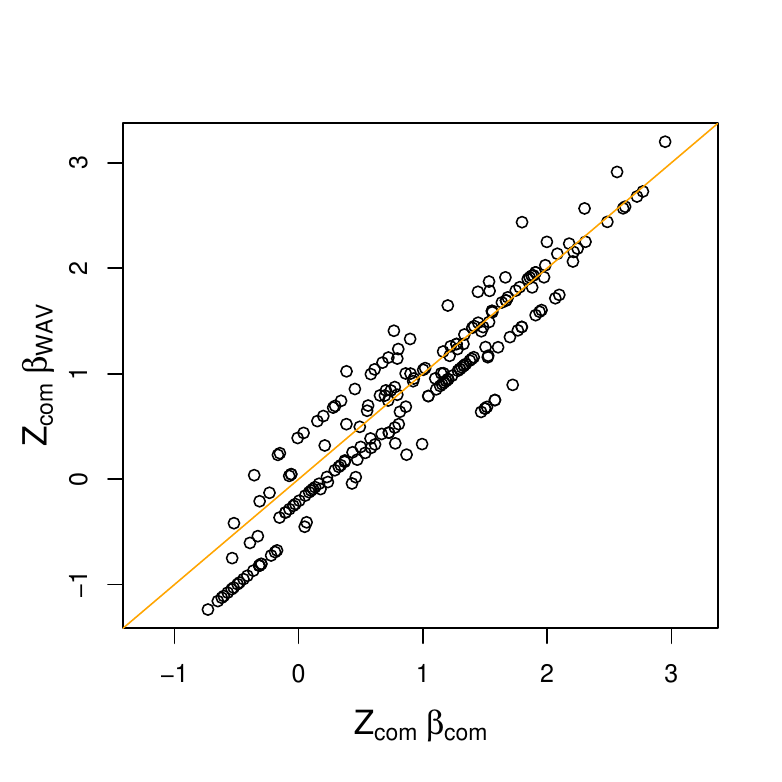}
\\[-8pt]
\includegraphics[scale=0.4]{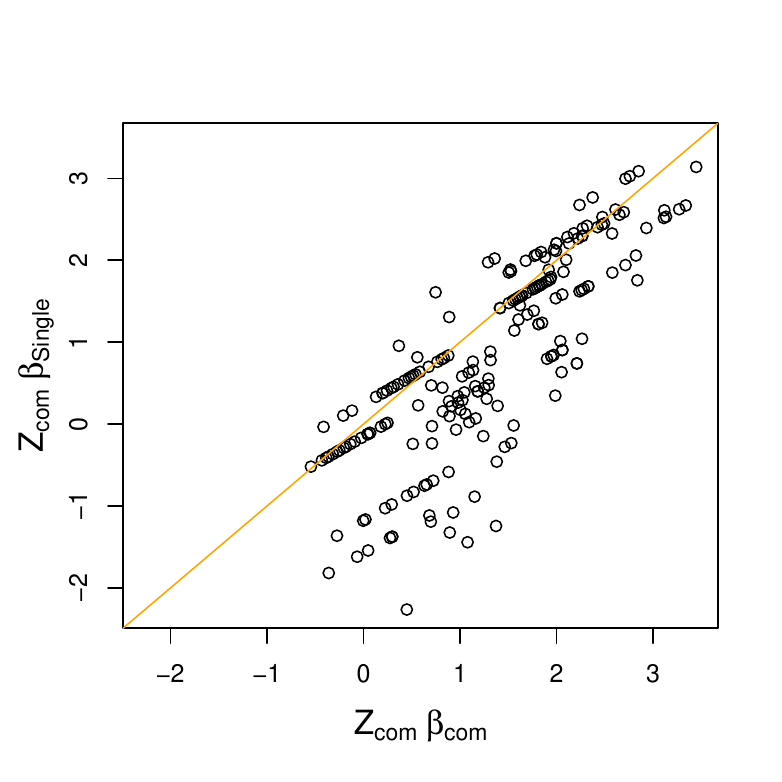}
\includegraphics[scale=0.4]{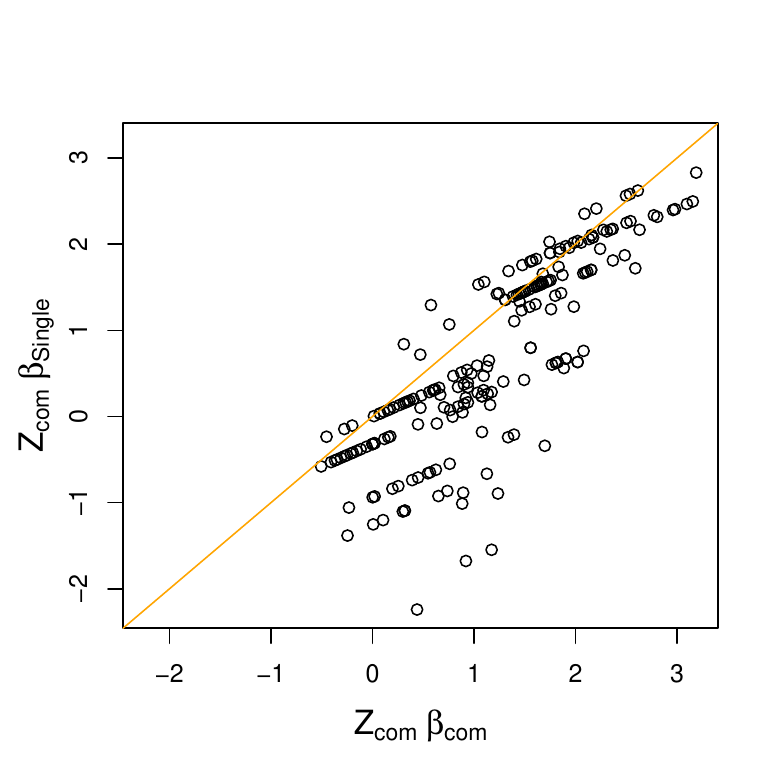}
\includegraphics[scale=0.4]{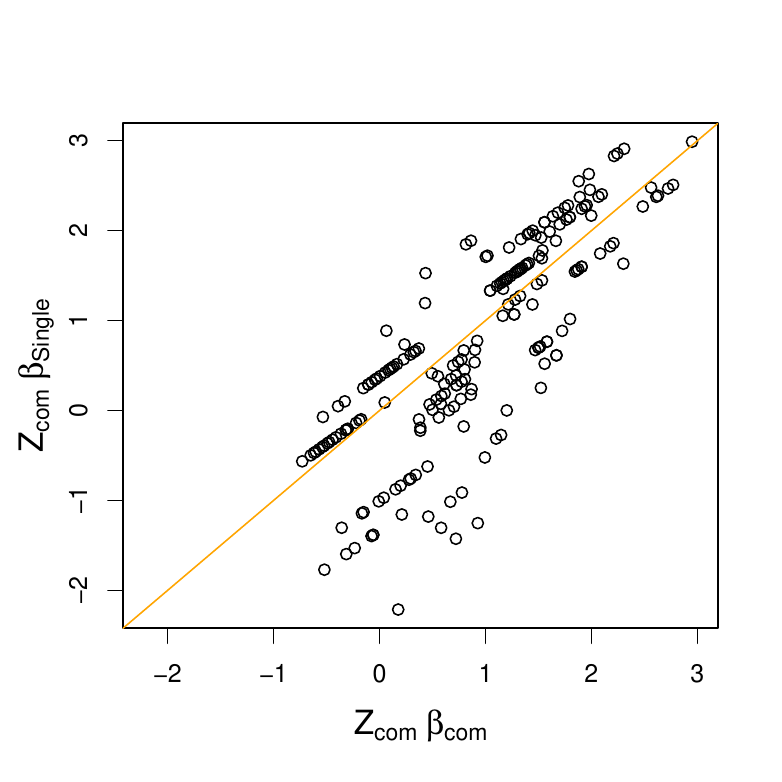}
\\[-8pt]
\includegraphics[scale=0.4]{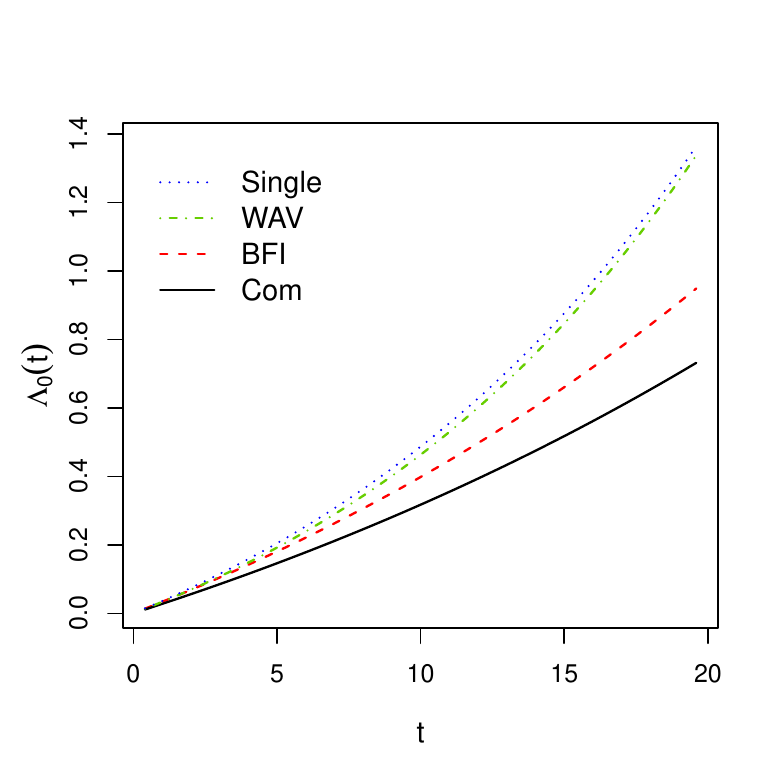}
\includegraphics[scale=0.4]{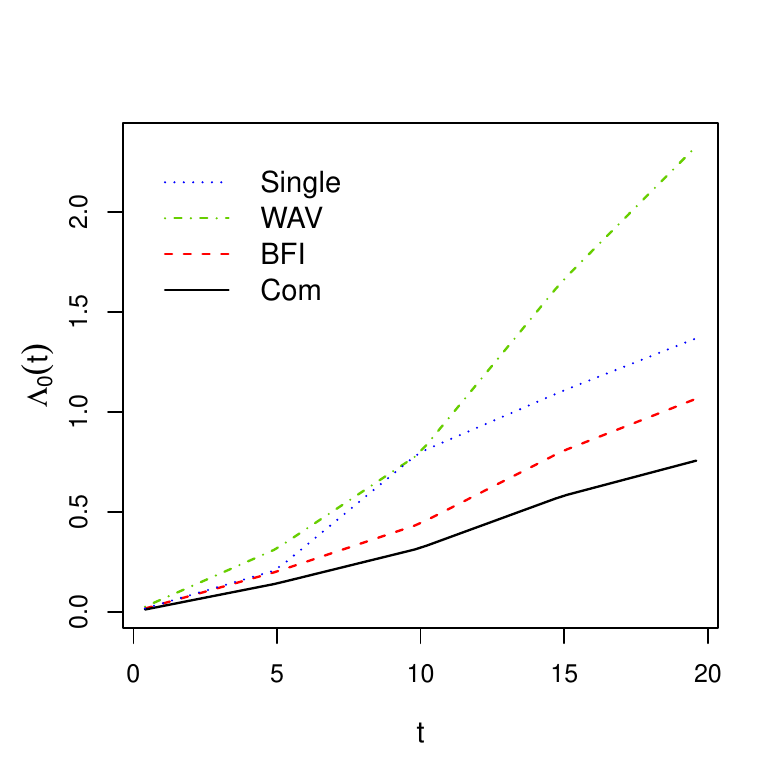}
\includegraphics[scale=0.4]{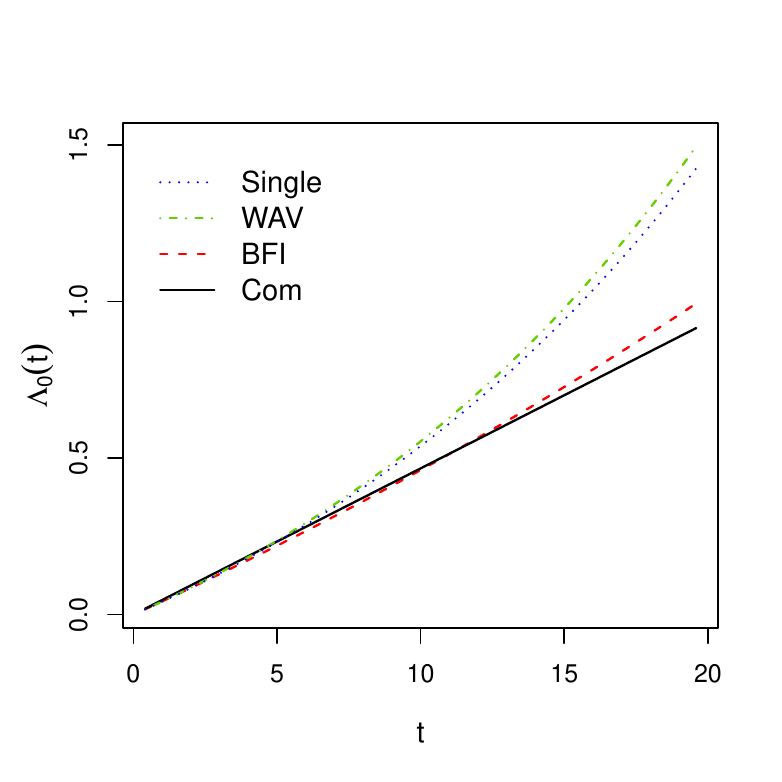}
\caption{Data from four hospitals. Scatter plots of $\bz^\top \widehat{\bbeta}_{BFI}$, $\bz^\top \widehat{\bbeta}_{WAV}$ and $\bz^\top \widehat{\bbeta}_{Single}$ against $\bz^\top \widehat{\bbeta}_{Com}$ in the first, second and third row, respectively, for 3 models: Gompertz (first column), piecewise constant with four intervals (second column), exponentiated polynomial (third column). Fourth row: estimates of $\Lambda_0$ in the three models: Gompertz (first plot), piecewise constant (second plot), exponentiated polynomial (third plot). The priors equal zero mean Gaussian distributions with diagonal inverse covariance matrices with $\gamma=0.1$ on the diagonal.}
\label{fig:4hospital}
\end{figure}

In this appendix the results of the simulation studies and the data analyses are presented. In the Tables \ref{tab:mseweib1} and \ref{tab:mseweib2} the $MSE$ for the regression parameters, their standard deviations, and the cumulative baseline hazard function in a series of time points are given, for multiple combinations of the sample sizes in three hospitals, different analysis models and the three estimators $\widehat{\bbeta}_{BFI}, \widehat{\bbeta}_{WAV}$ and $\widehat{\bbeta}_{Single}$. The data were simulated from a Weibull model. 
In Figure \ref{fig:3hospital} the performance of the BFI methodology is compared to the other proposed strategies. All salivary gland cancer patients had been randomly divided into three groups (i.e., three hospitals). 
In Figure \ref{fig:4hospital} similar plots are shown, but this time the estimates are based on the data of the patients from the four largest medical centers.

\begin{table}
	\centering
	\caption{Simulation results for $MSE_{\bbeta, BFI}$, $MSE_{\bbeta, WAV}$ and $MSE_{\bbeta, Single}$. All MSE values in this table have been multiplied by $10^{3}$ for clarity.
	\label{tab:mseweib1}}
 \scalebox{0.72}{
\begin{tabular}{ccccccccccccccccc}
		\toprule
		& && \multicolumn{4}{c}{$10^3 \times MSE_{\bbeta, BFI}$} && \multicolumn{4}{c}{$10^3 \times MSE_{\bbeta, WAV}$} && \multicolumn{4}{c}{$10^3 \times MSE_{\bbeta, Single}$}  \\ \cmidrule{4-7}\cmidrule{9-12}\cmidrule{14-17}
		$(n_1,n_2,n_3)$ & model  && $\beta_1$ & \multicolumn{1}{c}{$\beta_2$} & $\beta_3$ & \multicolumn{1}{c}{$\beta_4$} && $\beta_1$ & \multicolumn{1}{c}{$\beta_2$} & $\beta_3$ & \multicolumn{1}{c}{$\beta_4$} && $\beta_1$ & \multicolumn{1}{c}{$\beta_2$} & $\beta_3$ & \multicolumn{1}{c}{$\beta_4$}  \\
  \toprule
	$(50,50,50)$	& Exp   &&0.00 & 0.01 & 0.01 & 0.01 &  & 0.20 & 0.21 & 0.23 & 0.22 &  & 4.41 & 5.11 & 4.25 & 5.24\\

 \cmidrule{4-17}
		& Gom && 1.98 & 1.58 & 1.51 & 3.55 &  & 5.56 & 5.13 & 5.84 & 6.48 &  & 44.1 & 35.4 & 50.9 & 58.3\\

		\cmidrule{4-17}
  & Wei && 0.35 & 0.37 & 0.26 & 0.39 &  & 5.16 & 4.44 & 5.41 & 4.53 &  & 40.4 & 32.2 & 30.0 & 39.0\\
  
		\cmidrule{4-17}
		& PW4 &  & 0.16 & 0.15 & 0.19 & 0.17 &  & 2.93 & 1.87 & 2.14 & 1.89 &  & 30.9 & 17.6 & 23.7 & 21.2\\

\cmidrule{4-17}
		&PW8&& 0.30 & 0.23 & 0.22 & 0.49 &  & 3.55 & 2.96 & 3.06 & 3.54 &  & 40.6 & 32.2 & 26.2 & 38.6\\

		\cmidrule{4-17}
		& Poly  & & 0.74 & 0.42 & 0.62 & 0.60 &  & 7.90 & 7.19 & 5.68 & 8.43 &  & 64.8 & 43.9 & 52.6 & 65.3\\

 \toprule
	$(50,50,100)$	& Exp  & & 0.00 & 0.00 & 0.01 & 0.01 &  & 0.14 & 0.08 & 0.09 & 0.10 &  & 1.03 & 1.21 & 1.64 & 1.75\\
 
\cmidrule{4-17}
		& Gom & & 0.88 & 0.75 & 0.72 & 1.28 &  & 4.01 & 2.94 & 3.24 & 3.59 &  & 13.2 & 11.6 & 14.3 & 11.5\\

\cmidrule{4-17}
		& Wei  & & 0.19 & 0.23 & 0.22 & 0.25 &  & 3.20 & 2.50 & 2.37 & 3.77 &  & 12.1 & 15.3 & 9.87 & 12.1\\

		\cmidrule{4-17}
		& PW4  & & 0.07 & 0.08 & 0.08 & 0.10 &  & 1.37 & 1.17 & 1.23 & 1.10 &  & 8.20 & 7.80 & 5.35 & 8.96\\

    \cmidrule{4-17}
		&PW8&& 0.19 & 0.17 & 0.17 & 0.14 &  & 2.59 & 1.86 & 1.98 & 1.84 &  & 9.26 & 9.92 & 10.3 & 7.07\\

		\cmidrule{4-17}
		& Poly  & & 0.43 & 0.33 & 0.36 & 0.37 &  & 5.33 & 2.81 & 1.92 & 5.13 &  & 20.3 & 11.9 & 9.59 & 14.4\\

 \toprule
	$(50,100,100)$	& Exp  & & 0.00 & 0.00 & 0.00 & 0.00 &  & 0.05 & 0.07 & 0.08 & 0.06 &  & 1.78 & 1.92 & 1.92 & 1.69\\

 \cmidrule{4-17}
		& Gom && 1.89 & 0.97 & 0.90 & 1.73 &  & 2.13 & 2.23 & 1.63 & 2.26 &  & 18.1 & 17.6 & 11.0 & 21.8\\

\cmidrule{4-17}
		& Wei  & & 0.14 & 0.08 & 0.09 & 0.07 &  & 2.23 & 1.31 & 1.21 & 2.33 &  & 16.5 & 13.1 & 16.4 & 16.4\\
	\cmidrule{4-17}
		& PW4  & & 0.05 & 0.04 & 0.05 & 0.04 &  & 0.61 & 0.48 & 0.62 & 0.74 &  & 7.47 & 6.57 & 7.15 & 6.92\\

\cmidrule{4-17}
		&PW8& & 0.09 & 0.09 & 0.11 & 0.09 &  & 1.13 & 0.83 & 0.95 & 1.14 &  & 13.7 & 8.72 & 10.9 & 10.9\\

		\cmidrule{4-17}
		& Poly  & & 0.25 & 0.22 & 0.27 & 0.20 &  & 2.64 & 1.79 & 2.06 & 1.83 &  & 17.3 & 13.4 & 18.4 & 15.2\\

 \toprule
  	$(100,100,100)$	& Exp  & &0.00 & 0.00 & 0.00 & 0.00 &  & 0.04 & 0.04 & 0.06 & 0.04 &  & 2.01 & 2.28 & 1.71 & 1.75\\
 \cmidrule{4-17}
		& Gom & & 0.63 & 0.31 & 0.33 & 0.63 &  & 1.08 & 1.35 & 1.18 & 1.39 &  & 19.6 & 18.2 & 15.8 & 17.9\\

\cmidrule{4-17}
		& Wei &  &0.08 & 0.09 & 0.05 & 0.06 &  & 1.54 & 0.83 & 0.99 & 0.94 &  & 19.1 & 17.1 & 14.2 & 17.2\\

		\cmidrule{4-17}
		& PW4  & &  0.03 & 0.03 & 0.04 & 0.03 &  & 0.47 & 0.39 & 0.34 & 0.34 &  & 10.1 & 7.89 & 7.88 & 8.42\\
\cmidrule{4-17}
		&PW8&& 0.07 & 0.04 & 0.07 & 0.10 &  & 0.67 & 0.75 & 0.75 & 1.14 &  & 9.99 & 11.9 & 13.4 & 15.1\\

		\cmidrule{4-17}
		& Poly  & & 0.12 & 0.11 & 0.10 & 0.11 &  & 1.33 & 0.99 & 1.30 & 1.47 &  & 17.2 & 18.4 & 15.3 & 20.1\\

 \toprule
  	$(50,50,500)$	& Exp  & & 0.00 & 0.00 & 0.00 & 0.00 &  & 0.02 & 0.02 & 0.01 & 0.02 &  & 0.08 & 0.10 & 0.09 & 0.12\\

\cmidrule{4-17}
		& Gom && 1.13 & 0.51 & 0.55 & 1.06 &  & 0.43 & 0.33 & 0.32 & 0.49 &  & 0.55 & 0.71 & 0.77 & 0.71\\

\cmidrule{4-17}
		& Wei  & & 0.03 & 0.04 & 0.03 & 0.04 &  & 0.41 & 0.39 & 0.26 & 0.38 &  & 0.61 & 0.65 & 0.68 & 0.57\\

		\cmidrule{4-17}
		& PW4   && 0.01 & 0.01 & 0.01 & 0.01 &  & 0.17 & 0.12 & 0.17 & 0.13 &  & 0.34 & 0.31 & 0.38 & 0.37\\

\cmidrule{4-17}
		&PW8&& 0.03 & 0.02 & 0.02 & 0.03 &  & 0.25 & 0.17 & 0.17 & 0.19 &  & 0.59 & 0.39 & 0.48 & 0.52\\

		\cmidrule{4-17}
		& Poly  & & 0.04 & 0.05 & 0.05 & 0.08 &  & 0.45 & 0.38 & 0.40 & 0.38 &  & 0.95 & 0.99 & 0.74 & 0.88\\

 \toprule
  	$(50,100,500)$	& Exp  & & 0.00 & 0.00 & 0.00 & 0.00 &  & 0.01 & 0.01 & 0.01 & 0.01 &  & 0.13 & 0.11 & 0.09 & 0.11\\

 \cmidrule{4-17}
		& Gom & & 0.72 & 0.29 & 0.31 & 0.82 &  & 0.34 & 0.29 & 0.23 & 0.36 &  & 0.94 & 1.13 & 1.12 & 1.17\\

\cmidrule{4-17}
		& Wei  & &  0.03 & 0.02 & 0.01 & 0.02 &  & 0.36 & 0.24 & 0.26 & 0.23 &  & 0.95 & 0.85 & 0.71 & 0.77\\

		\cmidrule{4-17}
		& PW4  & & 0.01 & 0.01 & 0.01 & 0.01 &  & 0.07 & 0.12 & 0.08 & 0.10 &  & 0.58 & 0.58 & 0.42 & 0.41\\

 \cmidrule{4-17}
		&PW8&& 0.02 & 0.02 & 0.02 & 0.02 &  & 0.17 & 0.15 & 0.16 & 0.18 &  & 0.61 & 0.65 & 0.55 & 0.57\\

		\cmidrule{4-17}
		& Poly  & & 0.04 & 0.02 & 0.02 & 0.04 &  & 0.27 & 0.20 & 0.22 & 0.30 &  & 1.02 & 0.74 & 0.84 & 0.85\\

 \toprule
  	$(100,100,500)$	& Exp  & & 0.00 & 0.00 & 0.00 & 0.00 &  & 0.01 & 0.01 & 0.01 & 0.01 &  & 0.15 & 0.13 & 0.12 & 0.15\\

 \cmidrule{4-17}
		& Gom &&  0.38 & 0.16 & 0.19 & 0.40 &  & 0.29 & 0.38 & 0.23 & 0.36 &  & 1.28 & 1.61 & 1.22 & 1.44\\

\cmidrule{4-17}
		& Wei  & & 0.01 & 0.01 & 0.02 & 0.02 &  & 0.25 & 0.26 & 0.15 & 0.25 &  & 1.13 & 0.87 & 1.15 & 0.99\\

  		\cmidrule{4-17}
		& PW4   && 0.01 & 0.00 & 0.01 & 0.00 &  & 0.07 & 0.08 & 0.07 & 0.07 &  & 0.65 & 0.56 & 0.65 & 0.57\\

 \cmidrule{4-17}
		&PW8&& 0.01 & 0.01 & 0.01 & 0.01 &  & 0.15 & 0.11 & 0.11 & 0.15 &  & 0.93 & 0.81 & 0.62 & 0.93\\

		\cmidrule{4-17}
		& Poly  & & 0.02 & 0.02 & 0.02 & 0.04 &  & 0.22 & 0.20 & 0.29 & 0.39 &  & 1.17 & 1.40 & 1.20 & 1.28\\

 \toprule
  $(500,500,500)$	& Exp  & & 0.00 & 0.00 & 0.00 & 0.00 &  & 0.00 & 0.00 & 0.00 & 0.00 &  & 0.38 & 0.35 & 0.32 & 0.34\\

 \cmidrule{4-17}
		& Gom &&  0.03 & 0.01 & 0.01 & 0.03 &  & 0.08 & 0.05 & 0.06 & 0.08 &  & 2.78 & 2.25 & 3.02 & 3.77\\

\cmidrule{4-17}
		& Wei  & & 0.00 & 0.00 & 0.00 & 0.00 &  & 0.04 & 0.02 & 0.03 & 0.03 &  & 2.13 & 1.47 & 1.83 & 1.72\\

		\cmidrule{4-17}
		& PW4  & &0.00 & 0.00 & 0.00 & 0.00 &  & 0.02 & 0.02 & 0.02 & 0.02 &  & 1.51 & 1.26 & 1.37 & 1.43\\

 \cmidrule{4-17}
		&PW8&& 0.00 & 0.00 & 0.00 & 0.00 &  & 0.03 & 0.03 & 0.03 & 0.04 &  & 2.91 & 2.14 & 2.34 & 2.37\\

		\cmidrule{4-17}
		& Poly  & & 0.01 & 0.00 & 0.00 & 0.00 &  & 0.06 & 0.05 & 0.05 & 0.05 &  & 2.76 & 2.28 & 3.07 & 2.90\\

		\bottomrule
	\end{tabular}
}
\end{table}

\begin{table}
	\centering
	\caption{Simulation results for $MSE_{\Lambda_0}(t^*)$ at four different quantiles, $MSE_{(\mathbf{M})_{kk}, BFI}$ and $MSE_{\bbeta, BFI, True}$ of the regression coefficients. To enhance clarity, all MSE values in the table have been multiplied by $10^{3}$, except for the $MSE_{(\mathbf{M})_{kk}, BFI}$ values, which have been multiplied by $10^{6}$.
	\label{tab:mseweib2}}
 \scalebox{0.7}{
	\begin{tabular}{ccccccccccccccccc}
		\toprule
		& & &\multicolumn{4}{c}{$10^3 \times MSE_{\Lambda_0}(t^*)$} && \multicolumn{4}{c}{$10^6 \times MSE_{(\mathbf{M})_{kk}, BFI}$} && \multicolumn{4}{c}{$10^3 \times MSE_{\bbeta, BFI, True}$}  \\ \cmidrule{4-7} \cmidrule{9-12}\cmidrule{14-17}
		$(n_1,n_2,n_3)$ & model  && $20\%$ & {$40\%$} & $60\%$ & {$80\%$} && $\beta_1$ & {$\beta_2$} & $\beta_3$ & {$\beta_4$} && $\beta_1$ & {$\beta_2$} & $\beta_3$ & {$\beta_4$}  \\
		\toprule
	$(50,50,50)$	& Exp   && 0.04 & 0.06 & 0.07 & 0.08 &  & 0.98 & 1.12 & 1.01 & 1.10 &  & 230 & 106 & 105 & 226\\
 \cmidrule{4-17}
		& Gom && 0.54 & 2.59 & 12.2 & 70.7 &  & 8.55 & 15.8 & 15.3 & 12.5 &  & 25.5 & 25.5 & 19.4 & 29.5\\

		\cmidrule{4-17}
		& Wei  & & 0.23 & 1.23 & 4.35 & 15.6 &  & 12.5 & 12.9 & 11.7 & 7.85 &  & 17.0 & 14.2 & 10.8 & 15.2\\

 \cmidrule{4-17}
		& PW4  & &  0.66 & 1.56 & 3.01 & 8.17 &  & 7.97 & 5.55 & 7.23 & 6.49 &  & 26.1 & 18.3 & 15.2 & 33.6\\

\cmidrule{4-17}
		&PW8&& 1.49 & 3.72 & 8.57 & 23.8 &  & 8.19 & 9.20 & 6.15 & 7.22 &  & 18.0 & 11.6 & 14.1 & 14.3\\

		\cmidrule{4-17}
		& Poly&   & 0.75 & 2.51 & 6.45 & 17.4 &  & 14.8 & 8.74 & 14.2 & 10.3 &  & 15.9 & 12.7 & 19.0 & 20.1\\

\toprule
	$(50,50,100)$	& Exp  & & 0.03 & 0.04 & 0.04 & 0.05 &  & 0.40 & 0.35 & 0.38 & 0.50 &  & 224 & 103 & 108 & 228\\

 \cmidrule{4-17}
		& Gom &&  0.24 & 1.01 & 4.35 & 23.8 &  & 5.40 & 5.68 & 5.84 & 6.35 &  & 22.3 & 19.0 & 16.8 & 19.3\\

\cmidrule{4-17}
		& Wei  & &0.17 & 0.85 & 2.90 & 10.1 &  & 4.30 & 4.94 & 3.62 & 4.30 &  & 7.43 & 8.81 & 10.1 & 11.9\\

	\cmidrule{4-17}
		& PW4  & & 0.33 & 0.80 & 1.50 & 3.64 &  & 2.89 & 3.20 & 2.90 & 2.39 &  & 26.9 & 14.1 & 13.6 & 26.8\\

\cmidrule{4-17}
		&PW8& &0.81 & 2.19 & 4.60 & 10.2 &  & 3.64 & 3.42 & 3.83 & 2.72 &  & 16.4 & 13.2 & 11.2 & 12.0\\

		\cmidrule{4-17}
		& Poly  & & 0.37 & 1.26 & 3.31 & 9.09 &  & 5.28 & 5.20 & 4.60 & 4.98 &  & 16.8 & 15.3 & 13.3 & 17.5\\

 \toprule
	$(50,100,100)$	& Exp  & & 0.02 & 0.02 & 0.03 & 0.03 &  & 0.27 & 0.23 & 0.27 & 0.20 &  & 227 & 98.2 & 102 & 228\\

 \cmidrule{4-17}
		& Gom && 0.23 & 0.71 & 3.41 & 22.4 &  & 3.37 & 2.94 & 4.58 & 2.75 &  & 18.0 & 11.0 & 12.3 & 17.2\\

\cmidrule{4-17}
		& Wei  & & 0.07 & 0.40 & 1.44 & 5.24 &  & 2.17 & 2.41 & 2.93 & 2.27 &  & 8.52 & 5.59 & 8.46 & 8.05\\
	\cmidrule{4-17}
		& PW4   && 0.18 & 0.46 & 0.83 & 1.90 &  & 1.36 & 1.45 & 1.47 & 1.21 &  & 25.5 & 15.3 & 13.8 & 23.4\\
 \cmidrule{4-17}
		&PW8&& 0.43 & 1.10 & 2.72 & 6.59 &  & 2.55 & 1.81 & 1.83 & 1.89 &  & 11.4 & 7.64 & 9.03 & 12.1\\

		\cmidrule{4-17}
		& Poly   && 0.20 & 0.71 & 1.93 & 5.57 &  & 2.16 & 2.59 & 3.94 & 2.96 &  & 8.28 & 8.91 & 8.20 & 8.24\\
\toprule
  	$(100,100,100)$	& Exp  & & 0.01 & 0.01 & 0.02 & 0.02 &  & 0.17 & 0.15 & 0.10 & 0.10 &  & 232 & 100 & 101 & 227\\
 \cmidrule{4-17}
		& Gom && 0.12 & 0.44 & 1.96 & 11.7 &  & 2.33 & 1.59 & 2.13 & 2.45 &  & 13.4 & 8.80 & 8.36 & 16.5\\

\cmidrule{4-17}
		& Wei  & & 0.06 & 0.30 & 1.03 & 3.44 &  & 1.63 & 1.53 & 1.63 & 1.80 &  & 6.93 & 6.10 & 6.23 & 7.78\\
\cmidrule{4-17}
		& PW4   && 0.15 & 0.35 & 0.61 & 1.32 &  & 0.74 & 0.98 & 0.61 & 0.89 &  & 20.8 & 11.7 & 13.2 & 21.0\\
 \cmidrule{4-17}
		&PW8&&  0.29 & 0.74 & 1.60 & 4.19 &  & 0.94 & 1.39 & 1.00 & 1.62 &  & 6.60 & 6.19 & 6.66 & 9.59\\

		\cmidrule{4-17}
		& Poly   && 0.11 & 0.42 & 1.18 & 3.54 &  & 1.85 & 2.01 & 1.57 & 1.91 &  & 12.6 & 9.14 & 7.85 & 8.87\\

 \toprule
  	$(50,50,500)$	& Exp   && 0.00 & 0.00 & 0.01 & 0.01 &  & 0.02 & 0.01 & 0.02 & 0.02 &  & 226 & 100 & 100 & 228\\

 \cmidrule{4-17}
		& Gom && 0.09 & 0.06 & 0.45 & 7.10 &  & 0.17 & 0.21 & 0.24 & 0.22 &  & 10.7 & 6.73 & 7.77 & 12.0\\
\cmidrule{4-17}
		& Wei  & &  0.02 & 0.09 & 0.30 & 0.98 &  & 0.13 & 0.22 & 0.16 & 0.15 &  & 3.31 & 3.35 & 2.40 & 2.78\\

	\cmidrule{4-17}
		& PW4  & & 0.04 & 0.10 & 0.18 & 0.40 &  & 0.10 & 0.11 & 0.12 & 0.11 &  & 25.8 & 11.3 & 12.2 & 26.3\\

 \cmidrule{4-17}
		&PW8&& 0.09 & 0.21 & 0.45 & 1.04 &  & 0.13 & 0.12 & 0.13 & 0.14 &  & 9.67 & 6.88 & 6.39 & 11.6\\

		\cmidrule{4-17}
		& Poly  & & 	0.04 & 0.12 & 0.33 & 0.94 &  & 0.29 & 0.28 & 0.23 & 0.42 &  & 6.34 & 4.27 & 3.23 & 6.21\\

 \toprule
  	$(50,100,500)$	& Exp  & & 0.00 & 0.00 & 0.00 & 0.00 &  & 0.01 & 0.01 & 0.02 & 0.02 &  & 227 & 101 & 101 & 226\\
\cmidrule{4-17}
		& Gom &&  0.07 & 0.09 & 0.41 & 4.84 &  & 0.19 & 0.24 & 0.15 & 0.31 &  & 8.71 & 5.43 & 4.47 & 9.21\\

\cmidrule{4-17}
		& Wei  & &  0.01 & 0.06 & 0.21 & 0.72 &  & 0.20 & 0.14 & 0.10 & 0.13 &  & 2.92 & 2.26 & 2.95 & 2.70\\

		\cmidrule{4-17}
		& PW4 &  &  0.03 & 0.08 & 0.15 & 0.31 &  & 0.07 & 0.07 & 0.08 & 0.08 &  & 21.5 & 10.3 & 10.9 & 22.0\\

 \cmidrule{4-17}
		&PW8&& 0.07 & 0.18 & 0.38 & 0.92 &  & 0.09 & 0.11 & 0.11 & 0.10 &  & 9.26 & 4.99 & 4.67 & 9.60\\

		\cmidrule{4-17}
		& Poly  & & 0.03 & 0.10 & 0.28 & 0.79 &  & 0.15 & 0.18 & 0.16 & 0.17 &  & 6.19 & 4.47 & 4.65 & 7.18\\

\toprule
  	$(100,100,500)$	& Exp  & & 0.00 & 0.00 & 0.00 & 0.00 &  & 0.01 & 0.01 & 0.01 & 0.01 &  & 227 & 100 & 100 & 226\\

 \cmidrule{4-17}
		& Gom & & 0.04 & 0.05 & 0.27 & 2.68 &  & 0.19 & 0.22 & 0.20 & 0.21 &  & 8.38 & 4.71 & 5.62 & 7.79\\

\cmidrule{4-17}
		& Wei  & &  0.01 & 0.05 & 0.19 & 0.68 &  & 0.12 & 0.10 & 0.15 & 0.15 &  & 2.54 & 2.15 & 2.51 & 2.81\\

		\cmidrule{4-17}
		& PW4  & & 0.02 & 0.06 & 0.10 & 0.22 &  & 0.07 & 0.05 & 0.07 & 0.08 &  & 19.5 & 9.17 & 10.3 & 21.0\\

 \cmidrule{4-17}
		&PW8&& 0.05 & 0.13 & 0.30 & 0.68 &  & 0.11 & 0.09 & 0.07 & 0.09 &  & 6.00 & 3.10 & 3.70 & 6.22\\

		\cmidrule{4-17}
		& Poly  & & 0.02 & 0.08 & 0.23 & 0.71 &  & 0.15 & 0.21 & 0.17 & 0.28 &  & 5.91 & 3.66 & 4.46 & 5.36\\

 \toprule
  $(500,500,500)$	& Exp &  & 0.00 & 0.00 & 0.00 & 0.00 &  & 0.00 & 0.00 & 0.00 & 0.00 &  & 227 & 101 & 101 & 227\\

 \cmidrule{4-17}
		& Gom && 0.00 & 0.02 & 0.08 & 0.44 &  & 0.05 & 0.03 & 0.03 & 0.03 &  & 5.36 & 2.83 & 2.24 & 4.83\\

\cmidrule{4-17}
		& Wei  & &0.00 & 0.01 & 0.03 & 0.09 &  & 0.01 & 0.01 & 0.01 & 0.01 &  & 1.51 & 0.98 & 1.22 & 1.07\\

		\cmidrule{4-17}
		& PW4  & & 0.01 & 0.01 & 0.03 & 0.05 &  & 0.01 & 0.01 & 0.01 & 0.01 &  & 17.4 & 8.50 & 8.29 & 17.3\\

 \cmidrule{4-17}
		&PW8&& 0.01 & 0.03 & 0.06 & 0.15 &  & 0.01 & 0.01 & 0.01 & 0.01 &  & 3.21 & 2.17 & 1.81 & 3.59\\

		\cmidrule{4-17}
		& Poly  & & 0.00 & 0.02 & 0.05 & 0.16 &  & 0.03 & 0.02 & 0.04 & 0.03 &  & 3.21 & 2.39 & 2.78 & 4.33\\

		\bottomrule
	\end{tabular}
}
\end{table}


\newpage

\end{document}